\documentclass[conference]{IEEEtran}
\usepackage[utf8x]{inputenc}
\usepackage[pdftex]{graphicx}
\DeclareGraphicsExtensions{.pdf, .jpg}
\graphicspath{ {./figures/} }
\usepackage{url}
\usepackage[caption=false,font=footnotesize]{subfig}
\usepackage[cmex10]{amsmath}
\usepackage{amsfonts}
\usepackage{algorithmic}
\usepackage{float}
\usepackage{xspace}
\floatstyle{ruled}
\newfloat{algo}{thp}{lop}
\floatname{algo}{Algorithm}

\usepackage[usenames,dvipsnames,svgnames,table]{xcolor}
\newcommand{\PIBB}{\ensuremath{PI^{\mbox{\sf\tiny BB}}}\xspace}

\newcommand{\vtheta}{\ensuremath{\mathbf{\theta}}\xspace}
\newcommand{\mSigma}{\ensuremath{\mathbf{\Sigma}}\xspace}
\newcommand{\N}[2]{\ensuremath{\mathcal{N}(#1,#2)}\xspace}

\newcommand{\vowel}[1]{\ensuremath{\backslash \mbox{#1} \backslash}}

\definecolor{jbcolour}{rgb}{0.1,0.4,0.4}

\newif\ifdraft\drafttrue

\ifdraft

\newcommand\pyo[1]{{\color{red}}{\color{red}}{\footnotesize \color{red}[#1 - \textbf{PY}]}} %pyo for Pierre-Yves Oudeyer
\newcommand\jb[1]{{\color{jbcolour}}{\color{jbcolour}}{\footnotesize \color{jbcolour}[#1 - \textbf{Jules}]}} %JB for Jules Brochard
\newcommand\cmf[1]{{\color{blue}}{\color{blue}}{\footnotesize \color{blue}[#1 - \textbf{Cl\'ement}]}} %CMF for Clément Moulin-Frier
\newcommand\fs[1]{{\color{green}}{\color{green}}{\footnotesize \color{green}[#1 - \textbf{Freek}]}} %fs for Freek Stulp

\else

\newcommand\pyo[1]{}
\newcommand\jb[1]{}
\newcommand\cmf[1]{}
\newcommand\fs[1]{}

\fi
\definecolor{myred}{rgb}{0.8,0,0}
\definecolor{mygreen}{rgb}{0,0.6,0}
\definecolor{myblue}{rgb}{0,0,0.7}
\usepackage{pdfsync}
\title{Emergent Jaw Predominance in Vocal Development through Stochastic Optimization}

\author{
\centering
\IEEEauthorblockN{Cl\'ement Moulin-Frier\IEEEauthorrefmark{1} \IEEEauthorrefmark{3} \IEEEauthorrefmark{7}, Jules Brochard\IEEEauthorrefmark{1} \IEEEauthorrefmark{4} \IEEEauthorrefmark{7}, Freek Stulp\IEEEauthorrefmark{1} \IEEEauthorrefmark{2} \IEEEauthorrefmark{5}, Pierre-Yves Oudeyer\IEEEauthorrefmark{1} \IEEEauthorrefmark{6}}
\IEEEauthorblockA{
\\
\IEEEauthorrefmark{1}  Inria and Ensta ParisTech, France \\
\IEEEauthorrefmark{2} German Aerospace Center (DLR), Institute of Robotics and Mechatronics, Wessling, Germany \\
\IEEEauthorrefmark{3} clement.moulinfrier@gmail.com, now at the SPECS laboratory at the Universitat Pompeu Fabra, Barcelona, Spain \\
\IEEEauthorrefmark{4} brochard.jules@gmail.com, now at the Institut du Cerveau et de la Moelle épinière, Paris, France \\
\IEEEauthorrefmark{5} mail@freekstulp.net \\
\IEEEauthorrefmark{6} pierre-yves.oudeyer@inria.fr \\
\IEEEauthorrefmark{7} The two first authors contributed equally to the paper.
}
}

\begin{document}
\maketitle

\begin{abstract}
Infant vocal babbling strongly relies on jaw oscillations, especially at the stage of canonical babbling, which underlies the syllabic structure of world languages. In this paper, we propose, model and analyze an hypothesis to explain this predominance of the jaw in early babbling. 

This hypothesis states that general stochastic optimization principles, when applied to learning sensorimotor control, automatically generate ordered babbling stages with a predominant exploration of jaw movements in early stages. The reason is that those movements impact the auditory effects more than other articulators. 

In previous computational models, such general principles were shown to selectively freeze and free degrees of freedom in a model reproducing the proximo-distal development observed in infant arm reaching. The contribution of this paper is to show how, using the same methods, we are able to explain such patterns in vocal development. 

We present three experiments. The two first ones show that the recruitment order of articulators emerging from stochastic optimization depends on the target sound to be achieved but that on average the jaw is largely chosen as the first recruited articulator. The third experiment analyses in more detail how the emerging recruitment order is shaped by the dynamics of the optimization process.

\end{abstract}

%\todo{Avoid ``reaching for a vowel'' throughout the paper. Perhaps ``trying to achieve a vowel'' or ``vocalizing a vowel''}

\section{Introduction}

%\subsection{Theories of the origins of syllabic structures in infant development}

%In the course of early infant vocal development the vocal tract articulators are not used equally. A recruitment structure is displayed where the jaw seems to play a special role, in particular regarding an important developmental change around the age of 7 months called \emph{canonical babbling}~\cite{Oller2000}. Before this period, human infants produce non-speech vocalisations which do not display the syllabic structure specific to human speech. This particular structure appears with canonical babbling, where the infant suddenly and robustly produces jaw cycles coupled with phonation. This results in the first proto-syllables, as we often hear from young infants: \emph{``babababa''}. Although controversial~\cite{Oller2000}, this is often considered as the onset of speech learning. %\todo{Check if previous three sentences still correct.}
In the course of early infant vocal development the vocal tract articulators are not used equally \cite{Green2000,Green2002}. A recruitment structure is displayed where the jaw seems to play a special role \cite{macneilage2000}, paving the way towards an important developmental change around the age of 7 months called \emph{canonical babbling}~\cite{Oller2000}. Before this period, human infants produce non-speech vocalisations which do not display the syllabic structure specific to human speech. This particular structure appears with canonical babbling, where the infant suddenly and robustly alternate vowels and consonants in its first proto-syllables (typically \emph{``babababa''}). Although controversial~\cite{Oller2000}, this is often considered as the onset of speech learning. Some vocal articulators are more relevant than others for producing such proto-syllables, where the jaw and the lips seem to play a predominant role \cite{Green2000}. Experimental data shows that in the earliest stages of speech development the articulatory control of the jaw precedes the one of the lips, suggesting that this developmental pattern influences the pattern of speech sound acquisition \cite{Green2002}.  Moreover, the predominance of jaw movements has been considered as an important factor influencing both speech evolution \cite{macneilage98} and development \cite{macneilage2000}. 
%This paper aims at proposing an original hypothesis on the origin of jaw predominance in speech development by drawing an anology with the predominance of proximal articulators in arm control development and emphasizing the role of stochastic optimization in such a process through a computational model. 

A number of hypotheses, which we discuss in more detail in Section~\ref{sec_related_work_vocal}, have been proposed to explain the particular role of the jaw in speech development and in particular in canonical babbling; Warlaumont provides a recent review~\cite{warlaumont2015phonetics}. 
%revision
Note however that both issues (on the origins of jaw predominance and of canonical babbling) are two distinct ones in speech science, although they are obviously linked (see above). This paper focuses on modeling the emergence of jaw predominance from learning mechanisms. 
One set of hypotheses poses that the predominance of the jaw in vocalisation arises from other precursor behaviors, such as mastication and ingestion behaviors in the so-called \emph{Frame/Content} theory \cite{macneilage98}, or communicative orofacial gestures, such as lip-smacking in non-human primates \cite{ghazanfar2012cineradiography}. Other works consider that such a rythmic behavior is not specific to speech because arm babbling also appears around 6 months \cite{iverson2007relationship}. This suggests that these rhythmic patterns could be due to a general brain dynamics reorganization during the first year of life. Another line of work considers that rhythmic jaw movement can be the result of sensorimotor and social learning processes. Warlaumont \cite{Warlaumont_2012_ICDL,Warlaumont_2013_NN,Warlaumont_2013_ICDL} proposes computational models of syllabic structure emergence based on social or intrinsic reinforcement. 
Curiosity-driven self-exploration and imitation have also been considered as causal mechanisms to model the emergence of canonical babbling in speech acquisition~\cite{Moulin-Frier_Frontiers_2013}.

%The aim of this paper is to propose and computationally support an original hypothesis regarding the predominance of jaw movements in infant speech development.
Such developmental organization observed in vocal learning has analogs in the developmental organization of the acquisition of other motor skills, such as infant arm reaching which is characterized by the proximo-distal law: when learning to reach visual targets with their hands, infants first explore movements of its proximal joints (shoulder) before exploring more distal joints \cite{Berthier99}. 
In previous work, we have shown that basic stochastic optimization principles allow the reproduction of this proximodistal ordering~\cite{stulp12emergent,stulp13adaptive}. This was explained by the fact that stochastic optimization drives automatically the learner to explore more the degrees of freedom which have maximal impact on the sensory effect at any given point of the learning process, and that these ``maximal impact'' degrees of freedom were changing as the learning process got the system closer to sensory goals. 

The contribution of this paper is to show that the same principles are also able to explain the predominance of the jaw in vocal development. Thus, the hypothesis we propose is that the predominance of the jaw may be (partially) explained by fundamental optimization principles applied in the vocal sensorimotor space. 

Our model uses stochastic optimization to reach a number of auditory goals (vocalising different vowels) by searching in the space of articulatory movements. The optimization and trajectory generation algorithms are the same as in ~\cite{stulp12emergent,stulp13adaptive}, except that trajectories now correspond to movements of a simulated vocal tract, rather than a simulated arm. As a side effect of this process, the available degrees of freedom are autonomously frozen or released according to which articulator is the most impactful with respect to the optimization process. For the development of reaching, we previously reproduced proximodistal ordering~\cite{stulp12emergent,stulp13adaptive}, indepedently of the goal to be reached. Analogously, we now observe that the jaw is in average the most used articulator, independent of the auditory goals, i.e. the vowel to be vocalised.

The advantage of our hypothesis is that it does not rely on phylogenetic precursors as in the Frame/Content theory. It also does not require curiosity-driven intrinsic motivations or social reinforcement (but either or both these mechanisms could model the processes which generate the auditory goals that are set by the experimenter in this article). The fact that canonical babbling arises so robustly across different groups of infants \cite{Oller2000}, suggests the existence of several driving forces \cite{warlaumont2015phonetics}. We do not suggest that our model is the \emph{only} force. It should rather be considered as a general, domain-independent force, complementary to domain-specific forces.

\section{Related Work}
\label{sec_related_work}

Understanding how sensorimotor learning is autonomously structured through developmental processes is a major goal of developmental sciences. Computational contributions in the field of developmental robotics have in particular focused on the role of exploration strategies in the structuring of learning processes.  Indeed, complex sensorimotor behavior acquisition by developmental agents, whatever their biological or mechanical/computational nature, implies dynamical interactions in a complex embodiment with the environment.
%Regardless of their biological or mechanical nature, the acquisition of complex sensorimotor behavior by learning agents implies a dynamical interactions between the embodied agent and its environment. Developmental robotics addresses this by studying the form and formation of learning structures. Learning structure serves here as framework for an agent to discover how to perform a task.
In such a context, pure random exploration does not provide adequate data to allow an efficient learning \cite{oudeyer2013intrinsically}. This is due to the high dimensionality of the involved sensorimotor spaces, the non-linearity and redundancy of the sensorimotor mappings and the significant cost in time and resources of performing informative sensorimotor interaction with the environment. These constraints force the agent to develop efficient exploration strategies, resulting in the formation of a structured learning schedule.

%Sensorimotor maturations are examples of such structures. Let us emphasizes two examples observed in infant development. The first one concerns arm control development, where an

The freezing and freeing of degrees of freedoms is an important family of structuring mechanisms, organizing efficiently learning and exploration in animals, and infants in particular \cite{Bernstein67,Berthier99}. In this section, we now discuss two domains, arm control and vocalisation development, in which these phenomena have been observed in infant development. 

\subsection{Development of Arm Control}

When learning to progressively control their arms, infants obey the so-called \emph{proximodistal law}. The infant first starts to learn how to control his shoulders, and later focuses on its elbows, its wrists and finally its fingers. A number of experimental studies characterized this phenomenon. For example, Berthier et al. \cite{Berthier99} showed that the development of early reaching in infants \cite{Bertenthal98} follows a proximodistal structure, where infants first learn to reach by freezing the elbow and the hand, while varying shoulder and trunk movements, and then progressively used more distal joints of the elbow and hand. Studies in adult motor skill acquisition showed similar patterning of freezing and freeing of degrees of freedom, applied to the acquisition or racketing skills \cite{southard1987changing}, soccer \cite{hodges2005changes} or skiing \cite{vereijken1992free}. %\cmf{the few last sentences are directly taken from Stulp and Oudeyer, can be useful to change it a bit.}

%Although this law has biological reasons due to myelination, progressively impacting peripheral neural structures in a proximodistal manner,
Although neural myelinisation process may guide aspects of this developmental process by progressively impacting neural structures in a proximodistal manner,
it has been shown that the proximodistal law can also be explained by general learning and optimization mechanisms \cite{Schlesinger00,stulp12emergent,stulp13adaptive}. 
Our previous work has shown that a quite simple stochastic optimization process, allowing a progressive learning of reaching arm movements by the minimization of a cost function, can display an emergent ordering of degree-of-freedom recruitment~\cite{stulp12emergent,stulp13adaptive}. Without being precoded, such an optimization process naturally recruits in priority proximal joints because they induce a wider range of effects by allowing the movement of the entire arm (e.g. a wider range of reached hand positions) when compared to distal joints. 

Therefore, an organism exploring its own sensorimotor abilities has interest in recruiting proximal joints first because such a strategy allows him to rapidly make a reasonable approximation of the range of possible effects. It is actually not surprising that both biology, through the myelination process, and cognition, through exploration strategies, appear to converge to a similar solution (proximodistal release of degrees of freedom) for the same optimization problem: how to efficiently learn the control of a complex motor apparatus.

\subsection{Vocal Development}
\label{sec_related_work_vocal}

During vocal development, the involved articulators (the jaw, the tongue, the lips \ldots) are not recruited equally \cite{Green2000,Green2002}. In particular, at seven months canonical babbling appears in a very robust way \cite{Oller2000} and the predominance of jaw movements in vocal development has been suggested as an important factor influencing this phenomena \cite{macneilage2000}. 

The Frame/Content theory \cite{macneilage98,MacNeilage2001} suggests that jaw predominance is due to the role of feeding movement in speech evolution by providing a powerful sound modulation ability when coupled with phonation. Although controversial (see e.g. \cite{Arbib2005}), this hypothesis is supported by infant data displaying statistical vowel-consonant associations in line with the theory predictions.
A relatively similar hypothesis has been proposed where the considered phylogenetic precursor would be stereotyped communicative orofacial actions like lip-smacking in macaque monkeys \cite{ghazanfar2012cineradiography,ghazanfar2014evolution}. 

Sensorimotor and social learning processes have also been used to explain rhythmic jaw movement. Warlaumont \cite{Warlaumont_2012_ICDL,Warlaumont_2013_NN,Warlaumont_2013_ICDL} proposes computational models of syllabic structure emergence based on social or intrinsic reinforcement. The model starts with random vocalisations produced using an articulatory synthesizer, i.e a computer model of the human vocal tract able to synthesize sound from articulatory movements. These random vocalisations mainly result in non-speech sounds. In the first case (social reinforcement), a human subject then listens to these vocalisations and is asked to reinforce them or not according to his own judgment of speech-likeness. In the second case (intrinsic reinforcement), vocalisations are reinforced using an objective salience measure (more salient sounds are reinforced more). When reinforced, a learning rule drives the system towards producing the corresponding articulation more often. In both cases (social and intrinsic), the model converges towards the production of syllabic vocalisations.

Still based on sensorimotor and social learning, some of the authors of the present paper have proposed a model of speech acquisition based on curiosity-driven self-exploration and imitation \cite{Moulin-Frier_Frontiers_2013}. Here the agent drives its own vocalisations according to a learning-progress maximization principle: it self-generates auditory goals according to the progress it observes in learning how to achieve them. % while trying to reach them.
Such a mechanism has been shown to be highly efficient in learning inverse models in high-dimensional and redundant robotics setups  \cite{Oudeyer2007ITEC, Baranes2012RAS,Oudeyer2011Developmentalconstraintson, gottlieb2013information} and concretely formalizes concepts of intrinsic motivation described in the psychology literature into algorithmic architectures that can be experimented in computers and robots \cite{Schmidhuber91c,Barto04,Oudeyer2007FN,Baldassarre2011DLI2IIC}. A side effect of this exploration strategy is to self-organize developmental pathways where the agent autonomously focuses on tasks of increasing complexity. In particular, this leads to the emergence of canonical babbling where proto-syllabic structures are observed. 
%Applied to vocal learning, we observe that the model first produces articulations mainly resulting in no sounds and then more and more complex productions from unarticulated to proto-syllabic vocalisations.

In the following sections, we adapt the aforementioned model explaining the proximodistal law of arm development from stochastic optimization mechanisms \cite{stulp12emergent,stulp13adaptive} to the vocal domain. For this aim, we use a computer-simulated vocal tract model, and show that a similar effect can occur in particular conditions, recruiting the jaw in priority due to the wider range of auditory effects it produces, and displaying a coherent ordering of the other motor parameters.

\section{Methods}

This section first describes how vocal articulation and auditory perception are implemented in the proposed model. Then we define the cost function that expresses the aim of learning: to produce a vocalisation acoustically close to a target vowel. Finally we explain the stochastic optimization algorithm used to optimize this cost function.

%\subsection{Vocal tract: the Maeda model}
\subsection{Simulating Vocal Production and Perception}
\label{sec:simulation}

We now describe the articulatory synthesizer used, as well as how the motor trajectories that determine the vocal tract's shape over time are generated. We also describe how the output of the synthesizer -- peak frequency trajectories in Hertz -- is converted into a perceptual scale.

\subsubsection{Articulatory synthesis}

\begin{figure}[!t]
\centering
\includegraphics[width=3in]{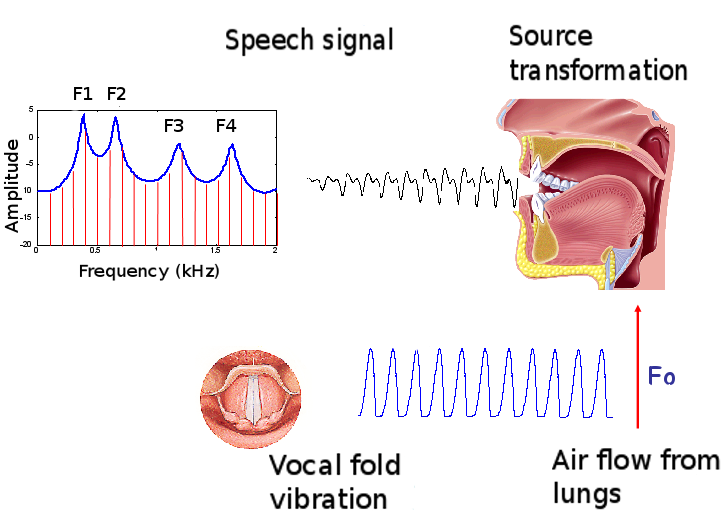}
\caption{An articulatory synthesizer is a model of human vocal production. The lung air flow triggers vocal fold vibration, providing a source signal with fundamental frequency $F0$. According to the vocal tract shape, acting as a resonator, the harmonics of the source fundamental frequency are selectively amplified or faded, resulting in a sound wave originating at the lips. Represented in a frequency-amplitude space, this sound wave typically displays a number of amplitude local maxima, called the formants $Fi$ and ordered from the lowest to the highest frequency. The formants are known to be important auditory features in human speech perception.}
\label{fig:trans_artic_acoust}
\end{figure}

Our computational model involves the articulatory synthesizer of the DIVA model described in~\cite{guenther2006neural}\footnote{\label{foot:divaurl} DIVA is available online at \url{http://www.bu.edu/speechlab/software/diva-source-code}. DIVA is a complete neurocomputational model of speech acquisition, in which we only use the synthesizer computing the articulatory-to-auditory function.}, based on Maeda's model~\cite{Maeda1989}.
%Without going into technical details, the model corresponds to a computational approximation
It is a computational approximation of the general speech production principles illustrated in \figurename~\ref{fig:trans_artic_acoust}.

\begin{figure}[!t]%figure1
\centerline{\includegraphics[width=3.3in]{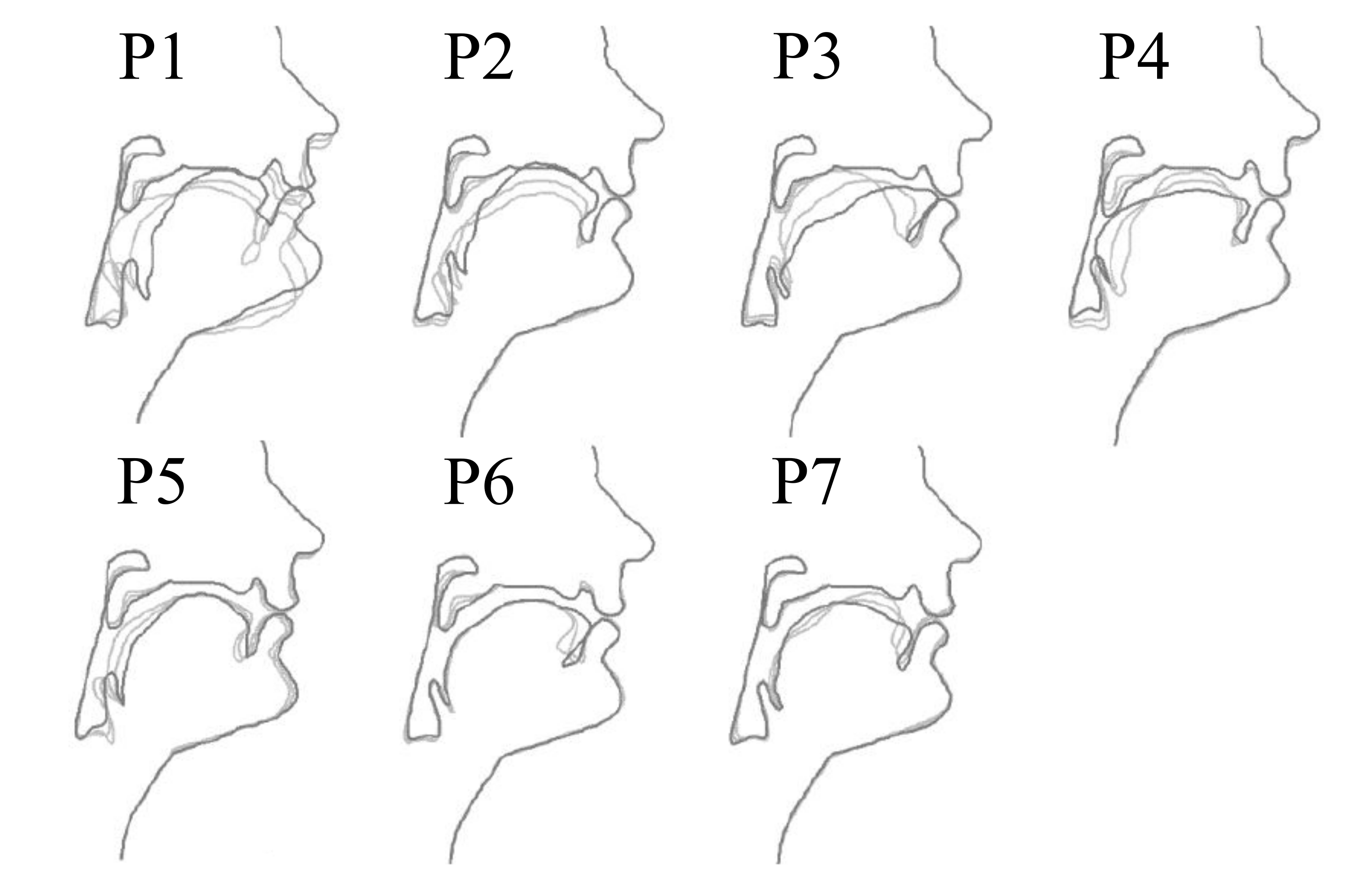}}% by writting ./ inclidegraphics is going to look for the figure file in the same folder that the *.txt file. If the figure is in an other directory change this
\caption{We use 7 articulatory dimensions to control the vocal tract shape (figure adapted from the DIVA source code documentation). Each subplot shows a sagittal contour of the vocal tract, where we can identify the nose and the lips on the right side. Bold contours correspond to a positive value of the articulatory parameter, the two thin contours are for a null (neutral position) and negative values. These dimensions globally correspond to the degrees of freedom of the human vocal tract articulators. For example, $P1$ mainly controls the jaw height, whereas $P3$ rather controls the tongue front-back position.   }
\label{fig:art_diva}
\end{figure}

The model %synthesizer
receives 13 articulatory parameters as input. The first 10 are extracted from a principal component analysis (PCA) performed on sagittal contours of images of the vocal tract of a human speaker, allowing the reconstruction of those contours from a 10-dimensional vector.
%In this study, we will only use the 7 first parameters (the effect of the others on the vocal tract shape is  negligible), fixing the 3 last in the neutral position (value $0$ in the software)
In this study, we only retain the 7 first parameters which  have a major impact on the shape of the vocal tract (labeled P1 to P7 on \figurename~\ref{fig:art_diva}). Since these parameters are derived from a statistical analysis of the vocal tract contours, they do not directly correspond to the biological speech articulators (the jaw, the lips, the tongue etc\ldots). However, the principal components extracted from the PCA relatively match the effect of vocal articulators on the vocal tract shape as it can be seen on \figurename~\ref{fig:art_diva}: P1 mainly correspond to jaw movements, P2 to tongue vertical movements, P3 to tongue horizontal movements, P4 to tongue flattening, P5 to low movements of the tongue at the level of the throat, P6 to horizontal movements of the tongue tip and P7 to small rotation movements of the tongue. The 3 last parameters of the 10-dimensional vector having a low influence on the vocal tract are fixed to their neutral position.
The effect of these 7 retained articulatory parameters on the vocal tract shape are displayed \figurename~\ref{fig:art_diva}. Through an area function, associating sections of the vocal tract with their respective areas for a given motor configuration, the model is able to compute the acoustic properties of the resulted signal if phonation occurs. This latter is controlled through 3 others parameters (glottal pressure, vocal cord tension and voicing), that we set at a value assuring normal phonation. We restrain our study to vowel sounds implying a sufficiently open configuration of the vocal tract. The synthesizer is then able to compute the 3 first formants (see \figurename~\ref{fig:trans_artic_acoust})  of the signal through the area function. 

\subsubsection{Motor trajectory generation}
\label{sec:motor_generation}

\begin{figure}[!t]
\centering
\includegraphics[width=3.5in]{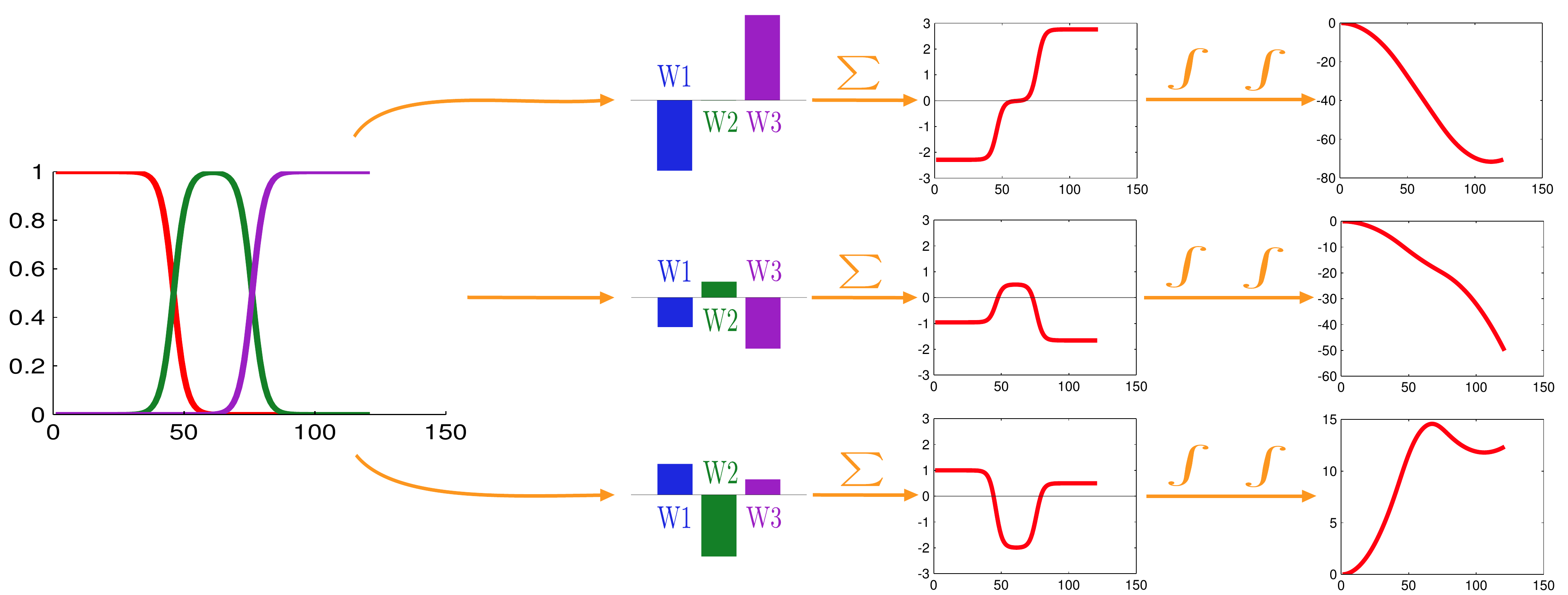}
\caption{Examples of basis functions' uses. Left: $3$ basis functions (x-axis is time). Center: $3$ triplets of weights and the weighted sum. Right: the $3$ position trajectories computed from the weighted sum of the basis functions, integrated twice.}
\label{fig:basis_functions}
\end{figure}

%A vocalisation corresponds to a trajectory of the 7 articulators displayed Figure~\ref{fig:art_diva} lasting $500 ms$. To do so, we encode the acceleration profile of each articulator as a linear combination of predefined basis functions. Starting from the neutral position of the vocal tract, these acceleration are integrated twice over the $500 ms$, thus resulting in position trajectories for each articulators, as illustrated on Figure~\ref{fig:basis_functions}.
A complete vocalisation corresponds to the time evolution of the vocal tract's shape. We constrain to vocalisations lasting $500 ms$ and control them through the acceleration of the 7 articulatory parameters. Taking inspiration from the Dynamic Movement Primitive framework \cite{ijspeert2002movement}, each acceleration profile is a linear combination of predefined basis functions. Starting from the neutral position of the vocal tract with a null velocity, these accelerations are integrated twice over the $500 ms$, thus resulting in position trajectories for each articulators, as illustrated on \figurename~\ref{fig:basis_functions}. 

The acceleration $\ddot{q}_{m,t}$ of the $m$-th articulator at time $t$ is determined as a linear combination of basis functions (Equation \ref{equ_acc}, \ref{equ_basisfunctions} and \ref{equ_kernel}), where $\theta_{m,b}$ is the weight of the $b^{\mbox{\tiny th}}$ basis function of the $m^{\mbox{\tiny th}}$ articulatory parameter:
\begin{align}
\ddot{q}_{m}(t) &= \sum_{b=1}^B g_b(t)\theta_{m,b}  & \mbox{Acc. of  $m^{\mbox{\tiny th}}$ articulatory} \label{equ_acc}\\
g_b(t) &= \frac{\Psi_b(t)}{\sum_{i=1}^B \Psi_i(t) }& \mbox{Normalized basis functions} \label{equ_basisfunctions}\\
\Psi_b(t) &= \exp \left(-(t-c_b)^2/w^2 \right)  & \mbox{Kernel} \label{equ_kernel}
\end{align}
The centers $c_{b=1\dots B}$ of the kernels $\Psi$ are spaced equidistantly in the $500 ms$ duration of the movement, and all have a width of $w=50 ms$. In our experiments we will use $B=4$ basis functions for each articulator. Using 7 articulators as stated above, a full motor command %\todo{I would argue that this is not a motor command.} 
is therefore a 28-dimensional vector.

\subsubsection{Auditory perception}
\begin{figure}[!t]
\centering
\includegraphics[width=2in]{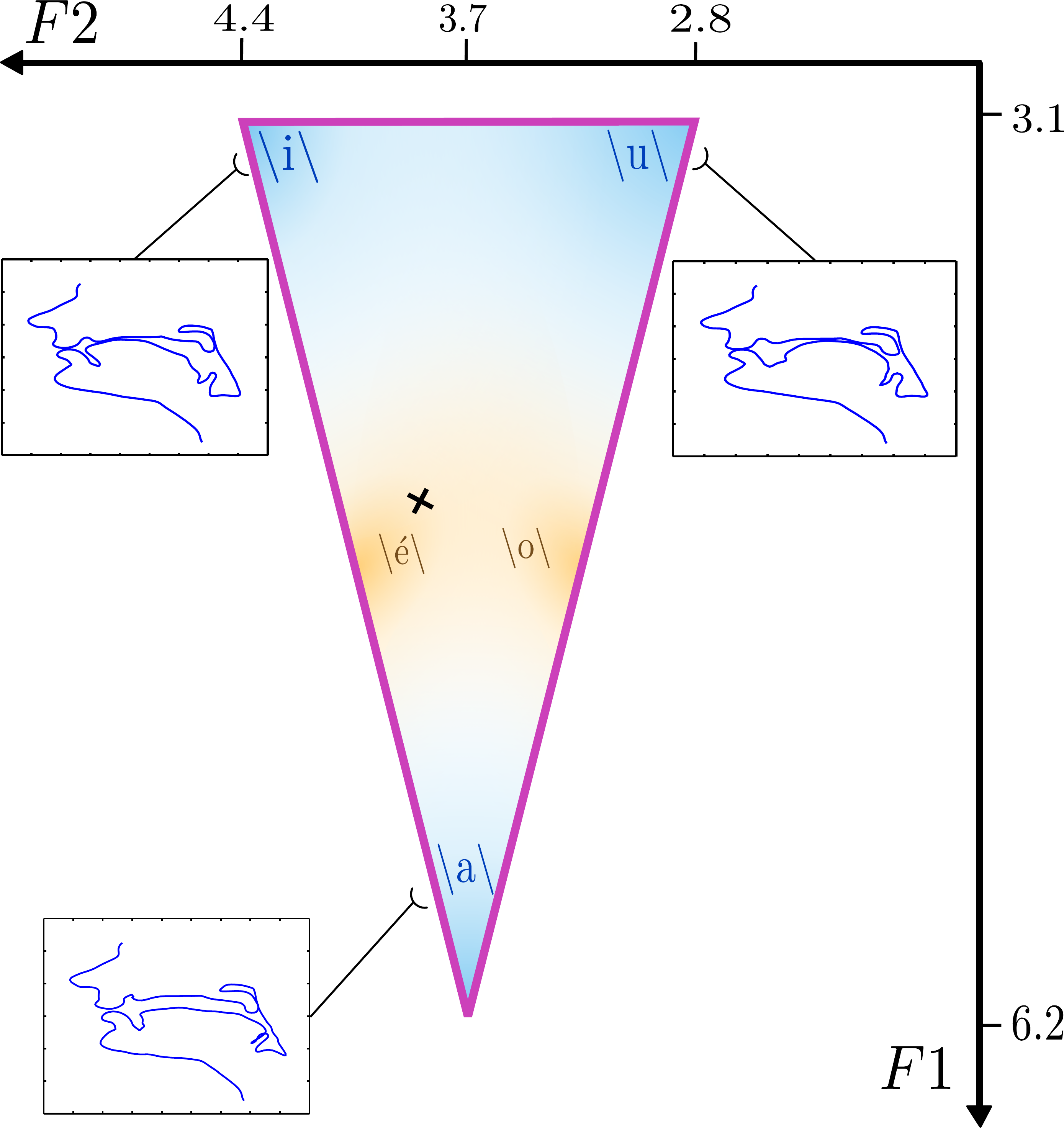}
\caption{\label{fig:vocalic_triangle} Variations of $F1$-$F2$ (in Barks scale): they take place in what is called the vocalic triangle, ideally represented here with the $5$ canonical vowels \vowel{a, e, i, o, u}. Vowels \vowel{i},\vowel{u} and \vowel{a} are shown in blue together with a possible tract shape for each of them. Vowels \vowel{e} and \vowel{o} are shown in brown. The cross represents the neutral vocalisation of the synthesizer, i.e. when all the articulators have a null value. Note that the $F1$-$F2$ axes are traditionally reversed in speech science to better reflect the relationship between formant values and articulatory configurations: high $F1$ values correspond to open configurations of the vocal tract (\emph{lowering} the jaw as in \vowel{a}, hence at the \emph{bottom} of the vocalic triangle), whereas high $F2$ values correspond to a constriction toward the front of the vocal tract as in \vowel{i}.  
%  \cmf{@Jules: What do the different colors inside the triangle represent? Is it feasible to have the true space generated by the synthesizer (as in Fig~\ref{simuSnapshots}) instead of a (too perfect) triangle? }
}

\end{figure}

The two first formants of the speech signal are known to be the main features of vowel perception \cite{klatt1982prediction,carlson1979vowel}.
The DIVA synthesizer provides formant trajectories as output, expressed in Hertz.  We convert the values in Hertz into a perceptual scale, typically linear at low frequencies and logarithmic at high frequencies. We use the Bark scale proposed by \cite{Schroeder1979} to reflect the psycho-acoustics of speech, defined as follow:
 \begin{equation}
Barks = 7\times sinh^{-1}\left(\frac{Hertz}{650}\right)
\label{hertzToBarks}
\end{equation}
 However, giving the same importance to all formants does not take into account spectral masking phenomena, according to which low-frequency components decrease the perceptual role of higher-frequency components. This led to the proposal \cite{Schwartz1997} that $F1$ should have typically three times the weight of an ``effective second formant'' $F'2$ grouping the roles of $F2$ and $F3$. Following this psycho-acoustic proposition, we modified the sensori space and used only the $(3F1,F2)$ plane. A similar modeling was used in \cite{moulinfrier2015cosmo} to successfully explain statistical tendencies in world-language phonological systems.
 %In fact, as the influence of $F3$ in the computation of $F'2$ appears to be unsignificant, we decide to approximate $F'2$ directly with $F2$.
Following this psychoacoustic proposition, we use $F1$ and $F2$ as auditory features and in the distance
computation in this $(F1, F2)$ plane we will weight the first dimension with a factor 3.
The perception that the vocal agent has of its own vocalisation is the value in Barks of $F1$ and $F2$ at the end of the vocalisation.
\figurename~\ref{fig:vocalic_triangle} displays possible values and associated vocal tract shapes.

%These constrains of the sensori-space restrict the audidatory space, which now cannot represent every human vocalisations. Nevertheless, it still captures the vowels vocalisations and achieve a sufficient balance between simplicity and descriptive power for our auditory experiments. All the vowels beeing characterize by the two first formants, our audidatory goals covers the early infant babbling. \jb{the 3 last sentences should probably be rephrased.}

 %The perception that the vocal agent has of its own vocalisation is the value in Barks of $F1$ and $F2$ at the end of the vocalisation.

\subsection{Cost function}
\label{sec:cost_function}

The main purpose of vocalizations in this model is to minimize the difference between a desired vocalization (set by the experimenter) and the actual vocalization (starting from an initial vocalization corresponding to the resting state of the articulators). We express this as a cost function that includes the distance to the goal in the formant-space: $ \left\| s_g - s \right\|^2 $, where  $s_g = \left( \begin{array}{c} 3F1_g \\ F2_g \end{array} \right)$ is the goal in the sensory formant space and
$s = \left( \begin{array}{c} 3F1_{t_N} \\ F2_{t_N} \end{array} \right)$ is formant values actually reached at the end of the vocalisation. Note the factor $3$ introduced for $F1$ values that models the higher weight of this formant as argued above. 
%\todo{MAJOR THING TO CLARIFY: in the previous paragraph it was said that F1 is weighted 3 times more than F2. Here the equations do NOT show this weighting. Where is the bug?}

Secondary costs that should also be minimized are the energy consumption, and awkward positions far from the rest position. The first is expressed as a penalty on accelerations: $\sum_{m=1}^7 \sum_{t=1}^{T}\frac{a_{m,t}^2}{2}$, where $a_{m,t}$ is the acceleration of the $m^{th}$ articulator at the time step $t$ of the vocalisation. To avoid pathological positions which are far from the resting position we add the term $\max_m\left(|P_{m,t_N}|\right)$, where $P_{m,t_N}$ is the position of the $m^{th}$ articulator at the end of the vocalisation. 

Weighting each part according to their impact on the cost, we get the function: 

\begin{multline}
J = 10^4 \underbrace{\left\| \left( \begin{array}{c} 3F1_g \\ F2_g \end{array} \right) - \left( \begin{array}{c} 3F1_{t_N} \\ F2_{t_N} \end{array} \right) \right\|^2}_{\mbox{Distance to goal}} \\ 
+ \underbrace{\max_m\left(|P_{m,t_N}|\right)}_{\mbox{\scriptsize Avoid awkward positions}} 
+ \underbrace{10^{-1}\sum_{m=1}^7 \sum_{t=1}^{T}\frac{a_{m,t}^2}{2}}_{\mbox{\scriptsize Penalize accelerations}}
\label{J}
\end{multline}.

A low-cost vocalisation is therefore a configuration which approaches the goal with a simple configuration and minimal energy. Note that this cost function is very similar to the one used to learn arm reaching behaviors~\cite{stulp13adaptive}.
The factors $10^4$ and $10^{-1}$ have two purposes: 1)~a scaling factor to compensate for different range of values the different cost components have, 2)~a weighting factor enabling the prioritization of tasks. The order of priorities is: achieve a sound close to the target, achieve end-state comfort, minimize accelerations.

%, we adapted their cost function which modeled the proximodistal structure of the arm. It was composed of three parts designed to favors movements close to a given goal and penalizes those far from the resting position or involving a high energy cost. Thus, the first component is the 

\subsection{Stochastic optimization}
\label{sec:stochastic_optimization}

The 28-dimensional vector that determines the movements of the vocal tract (see Section~\ref{sec:motor_generation}) defines a search space, whose optimum with respect to the cost function (Section~\ref{sec:cost_function}) defines a correct, low-energy vocalization of a particular vowel. In this section, we describe the stochastic optimization algorithm used to find this optimimum.

In principle, any stochastic optimization algorithm could be used, as long as it has the following properties:
1)~It is model-free, because we assume the learner does not have access to the details of the model described in Section~\ref{sec:simulation}. Such algorithms are known as black-box optimization algorithms, because they make no assumptions about the cost function, i.e. they treat it as a black box.
2)~Is able to adapt its exploration along different directions in the search space, because we want to study which parts of the vocal tract are explored most at different phases during learning.
3)~Does not contain any assumptions about vocal development.

Examples of algorithms that meet these requirements are natural gradient descent, cross-entropy methods or Covariance Matrix Adaptation -- Evolutionary Strategy (CMA-ES)~\cite{hansen2001completely}. We use a special case of CMA-ES, which is considered to be state-of-the-art in stochastic optimization, called \PIBB~\cite{stulp14simultaneous}. We do so because \PIBB is even simpler, and sets certain open parameters of CMA-ES to values that are derived from stochastic optimal control~\cite{stulp13robot}.

\begin{figure*}[!t]
\centering
\includegraphics[width=5in]{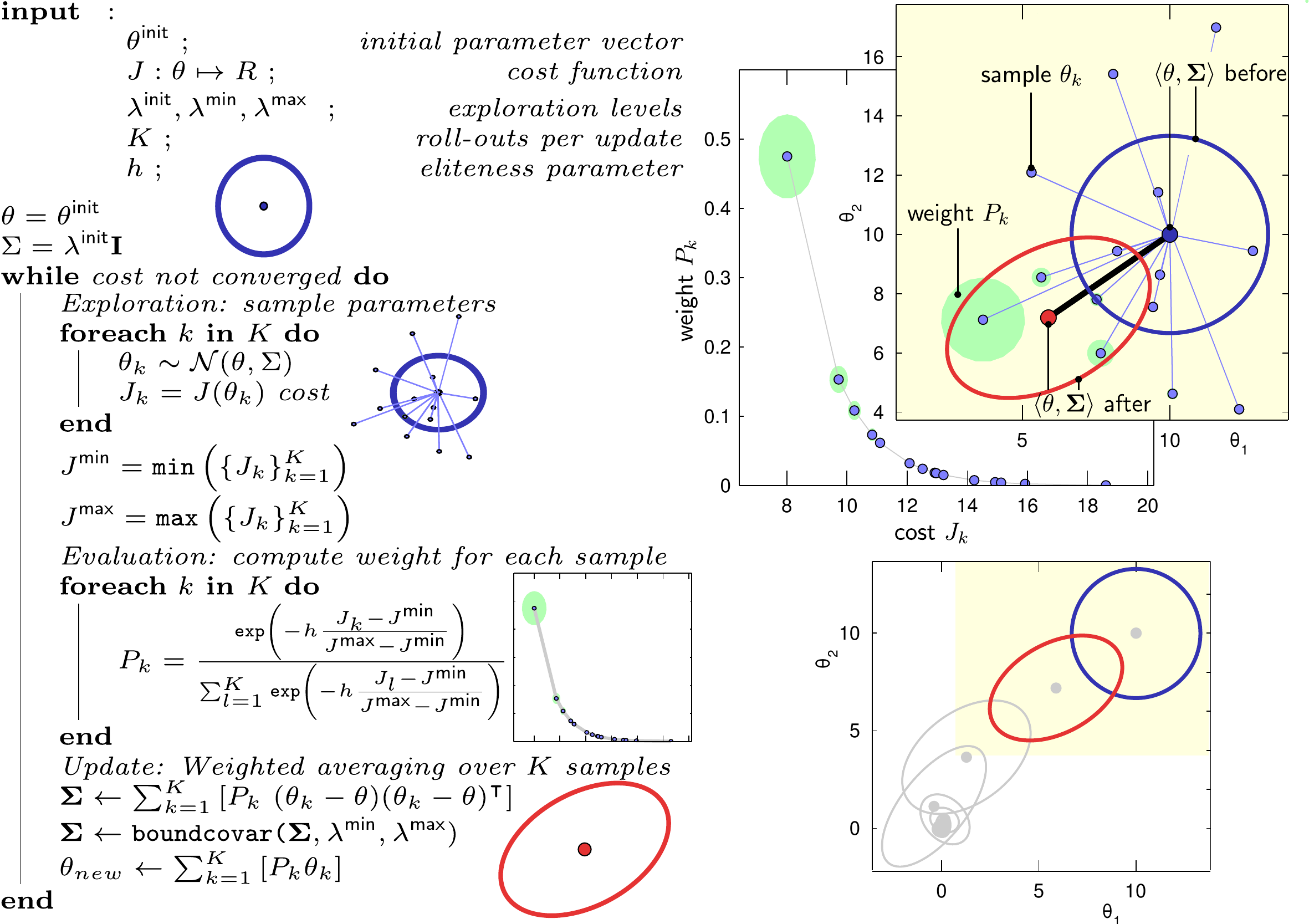}
\caption{Illustration of 
The \PIBB algorithm in pseudo-code (left), as well as a illustratory optimization (right). For simplicity, the search space is 2-dimensional and the cost of a sample \vtheta is simply the distance to the origin: $J(\theta) = ||\theta||$.  Top-right: Visualization of \emph{one} parameter update with \PIBB, and the decreasing exponential used to weight the samples. Bottom-right: Evolution of the parameters over \emph{several} updates, illustrating how the distribution converges towards the minimum $\theta_{g}=[0,0]$.
The algorithm is initialized by setting the mean and covariance parameters $\langle\theta,\Sigma\rangle$ to $\theta^{\mbox{\scriptsize init}}$ and $\lambda^{\mbox{\scriptsize init}}\mathbf{I}$ respectively, visualized as a dark blue ellipse. These parameters are updated at each iteration of the main loop. The red ellipse illustrates the first update.
 }
\label{fig:PIBB}
\end{figure*}

Within the search space, \PIBB represents a Gaussian distribution $\N{\vtheta}{\mSigma}$, where the aim is to find $\vtheta \equiv \vtheta_g$, where $\vtheta_g$ is the optimum. \PIBB is an iterative algorithm which consists of three phases during each iteration, which are visualized in \figurename~\ref{fig:PIBB}.

\begin{itemize}
\item \textbf{Exploration.} Sample $K$ parameter vectors $\vtheta_k$ from $\N{\vtheta}{\mSigma}$, and determine the cost $J_k$ of each sample. In the visualization of our illustratory example task $K=15$, and the cost $J(\vtheta)$ is the distance
$||\vtheta||$ to the center of the plane. In the \figurename~\ref{fig:PIBB} this distance between the blue dots and the center lies approximately between $8$ and $19$.

\item \textbf{Evaluation.} Determine the weight $P_k$ of each sample, given its cost. Essentially, low-cost samples have higher weights, and vice versa. The normalized exponentiation function that maps costs to weights is visualized in the top-right graph. Larger green circles correspond to higher weights.

\item \textbf{Update.} Update the parameters $\langle\vtheta,\Sigma\rangle$ with weighted averaging. In the visualization, the updated parameters are depicted in red. Because low-cost samples (e.g. a cost of 8-10) have higher weights, they contribute more to the update, and $\vtheta$ therefore moves in the direction of the optimum $\vtheta_{g}=[0,0]$.
\end{itemize}

%1)~sample $K$ parameter vectors from the current Gaussian distribution $\N{\vtheta_\mu}{\mSigma}$
%2)~compute the costs and corresponding weights for each sample
%3)~update the mean $\vtheta_\mu$ and covariance matrix \mSigma of the distribution.
%The details of the algorithm are explained in \figurename~\ref{fig:PIBB}. 

The bottom-right inset of \figurename~\ref{fig:PIBB} highlights two important properties of \PIBB, which it shares with cross-entropy methods and CMA-ES. First, the mean of the Gaussian distribution iteratively comes closer and closer to the optimum (the origin, in the example). Second, the covariance matrix of the distribution is elongated towards the optimimum.  \PIBB thus adapts its exploration, such that the amplitude of exploration is greater in the direction of the origin (i.e. the algorithm finds automatically that it is more efficient to try variations of $\vtheta$ along this direction at this stage of the optimization process). We define the {\it exploration magnitude} as the largest eigenvalue of the covariance matrix \mSigma. Intuitively, this means that the exploration magnitude is measured by the width of the gaussian in its largest direction. 
Below, this \PIBB optimization process is applied to the 7 groups of 4 parameters, each corresponding to one of the 7 articulators. Then, for each group the largest eigenvalue of each subset allows to measure the exploration magnitude of each articulator at any given point in the optimization process.

\subsubsection{Application of \PIBB to the synthesizer}

\begin{figure*}[!t]
\centering
\includegraphics[width=5in]{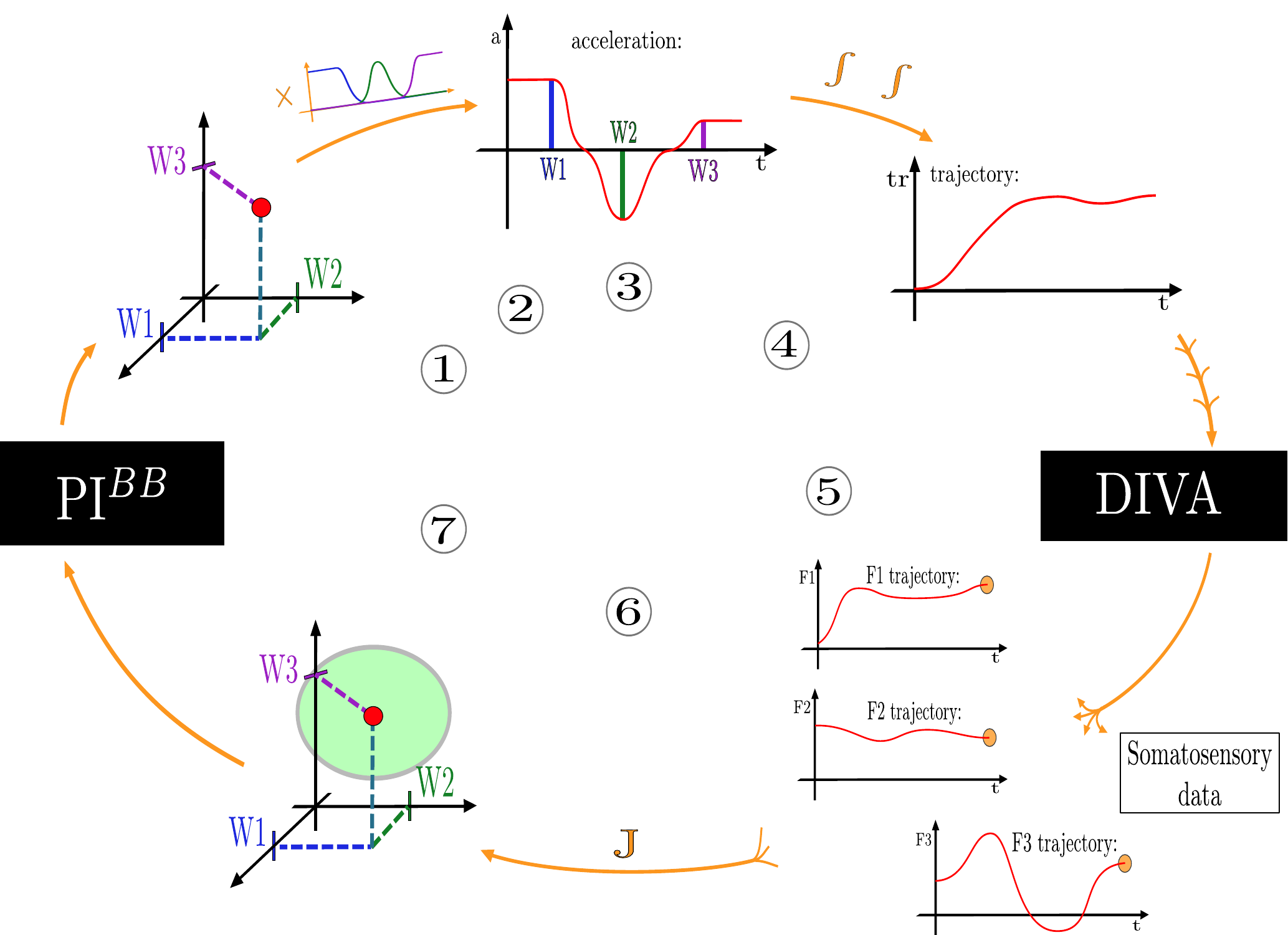}
\caption{\label{fig:simulation_loop} Illustration of applying \PIBB to vocalization learning.}
%To reach an auditory goal in the formant space, \PIBB is connected to the DIVA synthesizer. (1) to (4) are conducted for each motor parameter $Pi$ with $i=1..7$. (1) A point is drawn in a $B$-dimensional space, where B is the number of basis functions (here $B=3$ for the sake of illustration). (2) This point defines the weights of the $B$ basis functions. (3) The weighted sum of the 3 basis functions defines $Pi$ acceleration. (4) A double integration leads to the $Pi$'s position trajectory. (5) $P1$ to $P7$ trajectories are provided to the DIVA's synthesizer. (6) These trajectories and the corresponding auditory effect are used to compute the cost. (7) This cost is used to weight the sample of the $7$ $B$-dimensional points (i.e the same cost for each one).}
\end{figure*}

Before describing the experiments, \figurename~\ref{fig:simulation_loop} first summarizes how the vocal synthesizer, motor trajectory generation, auditory perception and stochastic optimization work together to learn how to vocalize vowels.

The following steps 1-6 are separetly performed $K=20$ times.
\begin{enumerate}
  \item For each of the $m=7$ vocal parameters, a 4D-vector is sampled from the distribution \N{\theta_m}{\mSigma_m}. Thus, one distribution is used per vocal parameter.
  
  \item Each of these 7 4D-vector defines the weights of the basis functions of its corresponding vocal parameters.
  
  \item The weights multiplied with the basis functions yield the accelerations for each of the 7 vocal parameters.
  
  \item Integrating these accelerations yields reference trajectories for the 7 vocal parameters.
  
  \item DIVA is then used to determine the vocalization (the formants in $Hz$), which are converted to auditory trajectories in the formant space.
  
  \item The cost for each of the sample is determined using the cost function $J$, see Equation (\ref{J}).

\end{enumerate}
Finally a step  7 is performed: the distribution \N{\theta_m}{\mSigma_m} is updated, by using the $K$ costs $J$ to weight the  $K$  4D-vectors. This is done separately for each of the 7 vocal parameters.

The steps above continue for 50 updates. After each update, we determine the exploration magnitude in vocal parameter $m$ by determining the largest eigenvalue  $\lambda_m$ of the covariance matrix $\mSigma_m$. Large values indicate that a lot of exploration is occuring within this vocal parameter.

\begin{figure}[!t]
\centering  
\subfloat[The relative exploration magnitude of each vocal parameter exploration during the optimization process are displayed by the seven color patches, from $P1$ (dark blue color patch at the bottom) to $P7$ (pink color patch at the top). Two exploration peaks are shown: one at update $13$ ($87\%$ relative exploration of $P1$) and one at update $20$ ($72\%$ of relative exploration of $P5$). The total exploration magnitude is plotted as a thick yellow/black line. Note that the values are smoothed to filter out the exploration peaks that only last one update: each value of the articulators' exploration is averaged with the $4$ last and next values.]{\includegraphics[width=3in]{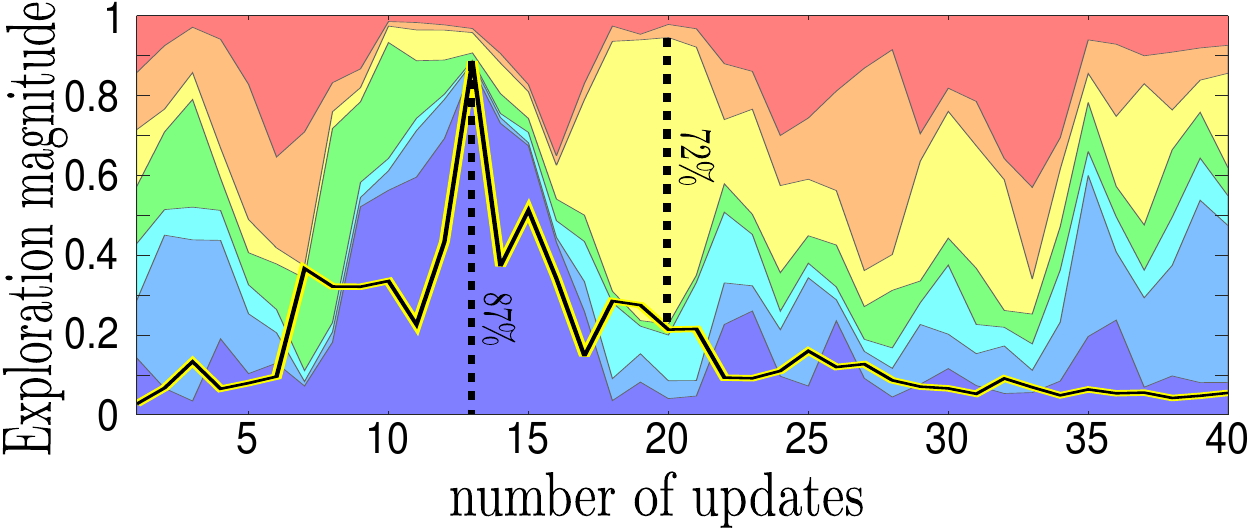}}

\subfloat[Learning curve of $J$ over the updates.]{\includegraphics[width=3in]{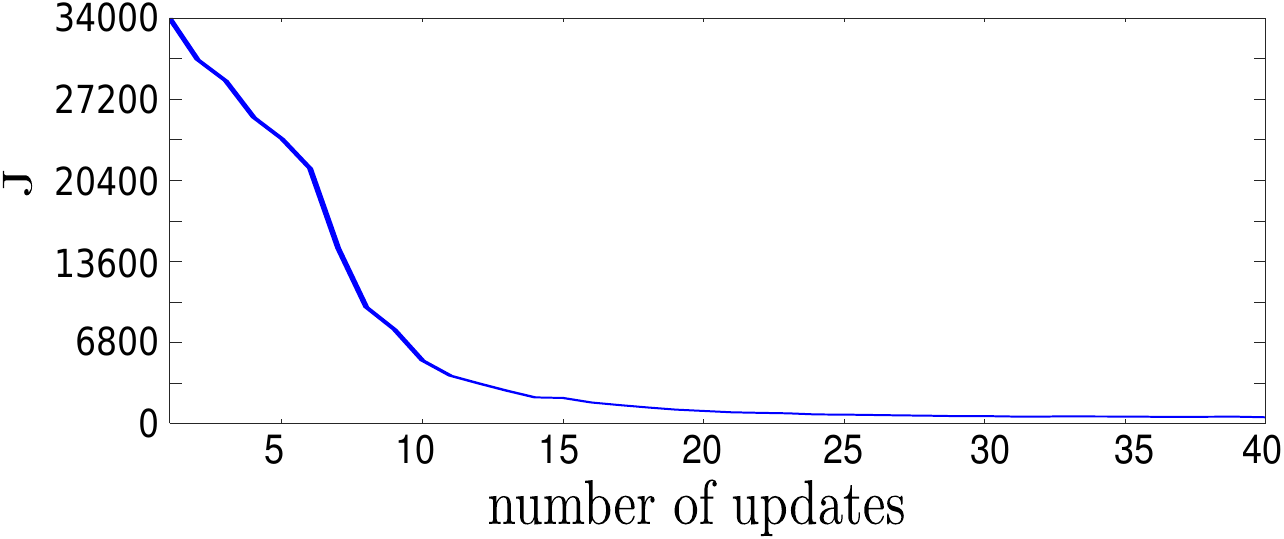}}
\caption{\label{fig:experiment_illustration} Illustration of stochastic optimization to learn to vocalize the vowel \vowel{a}. (a) displays the balance between the exploration of each articulators. (b) shows the optimization of the cost function $J$}
\end{figure}

\figurename~\ref{fig:experiment_illustration} illustrates one experiment, in which \PIBB learns to vocalize the vowel \vowel{a} within 50 updates. This graph illustrates the relative exploration magnitude within each of the 7 vocal parameters as learning progresses, i.e.  $\lambda_m^{\mbox{\tiny rel.}} = \frac{\lambda_m}{\sum_{i=1}^7\lambda_i}$. Thus, we see that initially the relative exploration magnitudes are the same, but between updates 5-20, most of the exploration happens in the first vocal parameter $P1$ (dark blue), which mostly corresponds to jaw movements. 

We interpret the relative exploration magnitude to be an indication of the relative freezing and freeing of degrees of freedom. For instance, the first vocal parameter $P1$ (dark blue, mostly controlling the jaw) is performing 89\% of the relative exploration at update 12. Thus, this degree of freedom is freed, whereas the others are frozen. At update 26, the $P5$ (back and forth tongue position) is freed, performing 83\% of the exploration.

In \figurename~\ref{fig:experiment_illustration}, the total exploration magnitude $\lambda^{\mbox{\tiny total}} =\sum_{m=1}^7\lambda_m$ is visualized as a thick yellow/black line. It is normalized to 1 within an experiment.

%\begin{figure}[ht]
%\includegraphics[width=0.9\textwidth]{experiment_illustration-svg}
%\caption{\label{fig:experiment_illustration} Illustration of stochastic optimization to learn to vocalize the vowel \vowel{a}. The relative exploration magnitude of each vocal parameter exploration is associated with a color. The total exploration magnitude is plotted as a thick yellow/black line.}
%\end{figure}

%To use the two previous model, we need a way to connect them. Our connection relies on motor trajectories generation and a cost function. The former allow us to go from \PIBB to the synthesizer, the later from the synthesizer to \PIBB.

%Given the vocal tract and auditory model and the stochastic optimization algorithm we just described, Figure \ref{fig:simulation_loop} illustrates the simulation loop. Each sample of \PIBB
%is composed of 7 real values,
%one for each articulatory parameters. Firstly, through basis functions (Figure \ref{fig:basis_functions}) and integration, each
%point
%array
%is transformed into a motor trajectory .
% These 7 motor trajectories represent a complete articulatory movement. From this latter, the vocal tract model then computes the corresponding auditory trajectories in the formant space. The cost of that vocalisation is computed from this articulatory and auditory trajectories, as we will explain more precisely when defining our experiments in the next section. Finally, a new Gaussian distribution is computed from these samples and their respective costs and the process repeats until convergence.

\section{Experiments}
\label{sec:experiments}

In this section, we first present experiments that investigate the freeing and freezing of degrees of freedom in the context of learning to vocalize the vowels \vowel{a}, 
\vowel{i}, and
\vowel{u}. Then, we do the same for arbitrary goals in the vocal space. Finally, we perform a sensitivity analysis providing further explanation of the obtained results.

\subsection{Experiment 1: Recruitment Order for Vowels as Goals}

This experiment analyses the recruitment order of parameter when learning how two achieve the three canonical vowels \vowel{a}, \vowel{i} and \vowel{u}. 

\subsubsection{Method and results}

\begin{figure}[!t]
\centering
\includegraphics[width=3.3in]{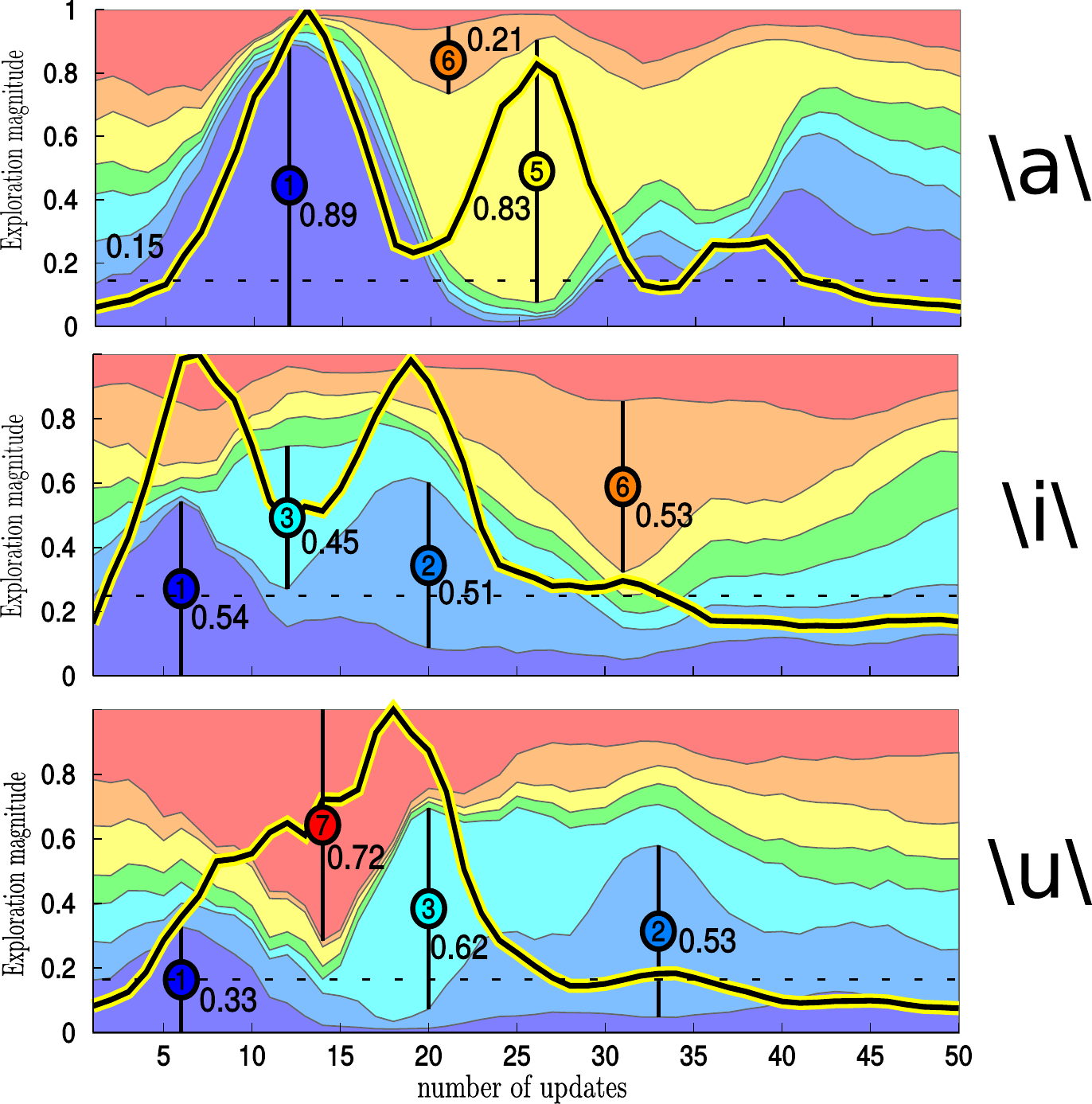}
\caption{\label{fig:results} Results of running stochastic optimization to learn to vocalize the vowels \vowel{a}, 
\vowel{i}, and
\vowel{u}. Same conventions as in \figurename~\ref{fig:experiment_illustration}.}
\end{figure}

We perform three optimization runs, one for each of the vowels \vowel{a}, \vowel{i} and \vowel{u}. The results are shown in \figurename~\ref{fig:results}, highlighting the order in which degrees of freedom are freed (a degree of freedom is said to be freed when it has a large relative exploration magnitude). We do so by annotating the graph with the maximum relative exploration, e.g. 0.89 for $P1$ in the first graph. This annotation is only added if the absolute normalized exploration (thick yellow/black line) is 5 percentage points above the baseline at the beginning; this threshold is indicated by a dashed line. Our motivation for this threshold is that, if there is no (absolute) exploration at all, none of the degrees are being freed, and it is therefore not relevant to study the relative exploration magnitudes. 
%\todo{@Jules: Explain smoothing. What was the window size? Why is this needed at all? (just tell me what the window size is otherwise and I'll adapt the text accordingly -- Clément)}

By analyzing the peaks of the maximum relative exploration, we can determine a {\it recruitment order}, i.e. the order in which the degrees of freedom are freed. For the vowel \vowel{a}, this order is $P1$, $P6$, $P5$. For the vowel \vowel{i} it is $P1$, $P3$, $P2$, $P6$, and for the vowel \vowel{u} it is $P1$, $P7$, $P3$, $P2$.

\subsubsection{Discussion}

%Extracting exploration magnitudes during the reaching attempts toward different goals in the vocalic triangle
%%allows
%enables us
%to analyze
%%in detail possible maturation phenomena in our model.
%the emerging recruitment stages of the degrees of freedom over the learning.
%Figure~\ref{recruitGoals} displays the result in two other particular simulations attempting to reach the three vowels /i/ and /u/ (a simulation for the vowel /a/ being already displayed on Figure~\ref{measureExplo}, bottom). Regarding the vowel /i/, we observe $4$ phases of recruitment. The $3$ first ones ($P1$,$P3$ and $P2$ maxima) appear together with a relatively high total exploration magnitude. The last phase ($P6$) is more an adjustment phase. Regarding the vowel /u/, we also observe $4$ phases of recruitment. The first and last ones ($P1$ and $P2$ maxima) appear together with a low total exploration magnitude and can therefore be considered as non relevant. The second and third ones ($P3$ and $P7$ maxima) appear together with a high total exploration magnitude.

% The goal is here the vowel \vowel{a}. This simulation shows $4$ phases of recruitment. The $3$ first ($P2$,$P1$ and $P6$) correspond to $3$ progressive improvements of the cost function. While the last phase ($P5$) is more an adjustment phase.

In these results we observe that the most recruited parameters are those controlling the distinctive phonetic features of the vowel to vocalize. Producing the vowel \vowel{a} necessitates a vocal tract constriction at the level of the throat, what is typically obtained by opening the jaw and placing the tongue in a back position, hence the recruitment of $P1$ (jaw opening) and $P5$ (tongue movement at the throat level). Producing \vowel{i} necessitates a constriction of the tongue behind the teeth, what is typically obtained with a rather closed jaw and the tongue in the front position, hence the recruitment of $P1$ (jaw), $P3$ (tongue horizontal movement) and $P2$ (tongue vertical movement). Finally, producing \vowel{u} necessitates constriction back in the palate, what is typically obtained by placing the tongue in a high-back position, which is here obtain through the conjoint action of $P1$ (jaw),
%$P1$ and $P7$ (CMF: not coherent, to check).
$P7$ (tongue rotation) and $P3$ (tongue horizontal movement).%\cmf{ not coherent, to check}

It therefore seems that attempting to achieve different vowel goals results in different recruitment orders. Moreover the recruited articulators for a given vowel are mainly those involved in the production of the associated phonetic features.

Another observation is that $P1$, the parameter related to jaw movements, is freed first for all the vowels. To investigate whether this is due to the choice of these particular vowels, or rather a general feature of learning vocalizations in this sensorimotor space, we conduct a second experiment.

\subsection{Experiment 2: Recruitment Order for Random Goals}

This experiment analyses the global tendency of the recruitment order on a large number of vowel goals uniformly distributed in the formant space. 

\subsubsection{Method and results}

We perform separate optimization runs for achieving 40~000 vowel goals drawn randomly in the vocalic triangle. For each run, we analyze the recruitment order using the same method as in the previous experiment. 

\begin{figure}
\centering
\includegraphics[width=3.3in]{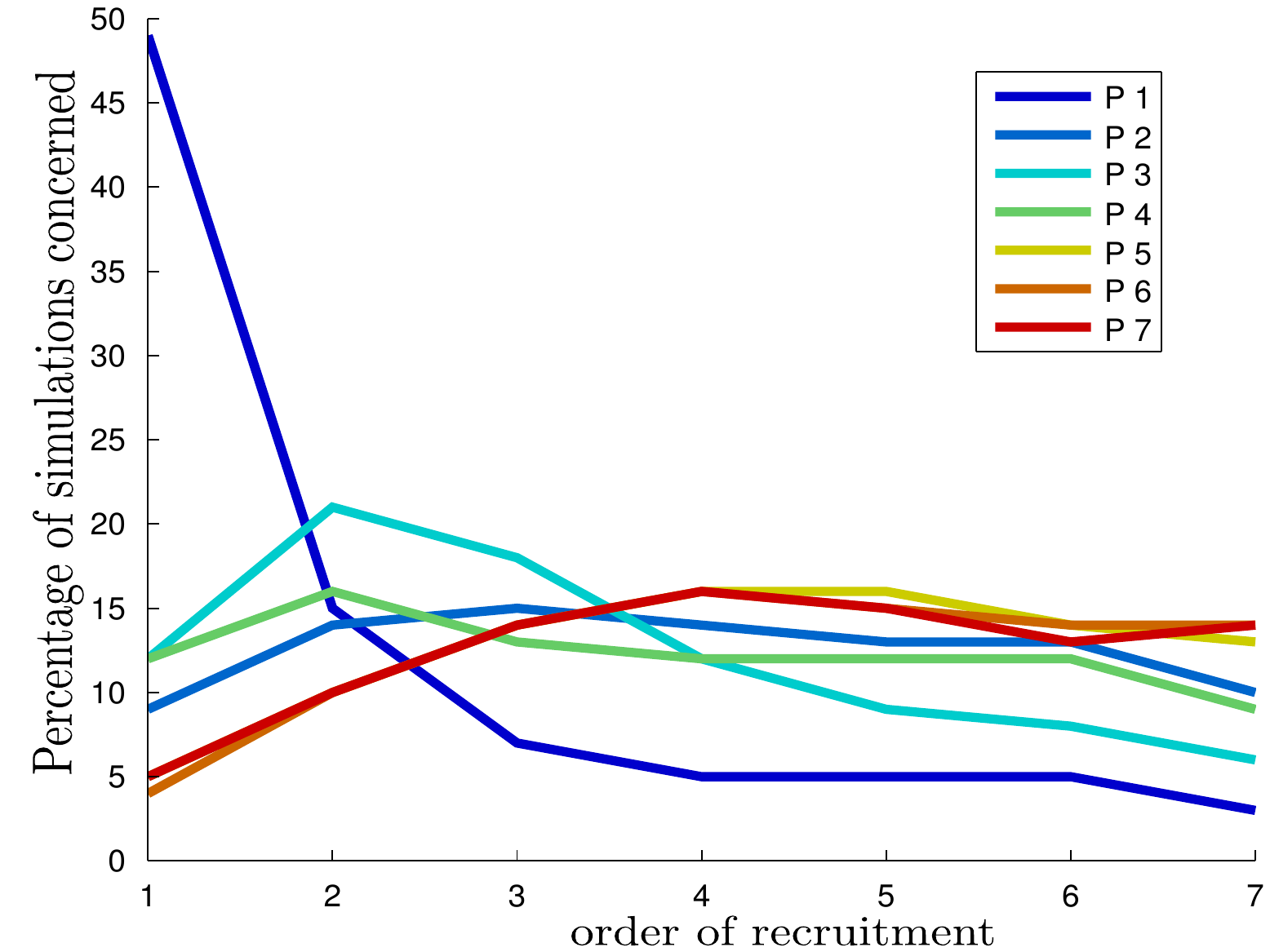}
%\caption{Frequencies of recruitment of each parameter over the whole vocalic triangle, and in each position.}
\caption{\label{globalRecrut} Rank frequencies for vowels uniformly distributed in the vocalic triangle. P1 is mostly related to jaw movements, see \figurename~\ref{fig:art_diva}.} % \jb{TODO: Unify notation with the previous figure (art1, art2, art3 vs P1, P2, P3)}}
\end{figure}

The results are shown in \figurename~\ref{globalRecrut}. We observe a strong tendency to recruit $P1$ first: it is the case in almost 50\% of the simulations, whereas the other articulators are recruited first only between 4 and 13\% of the simulations. 

\subsubsection{Discussion}

\begin{figure}
\centering
\includegraphics[width=3.7in]{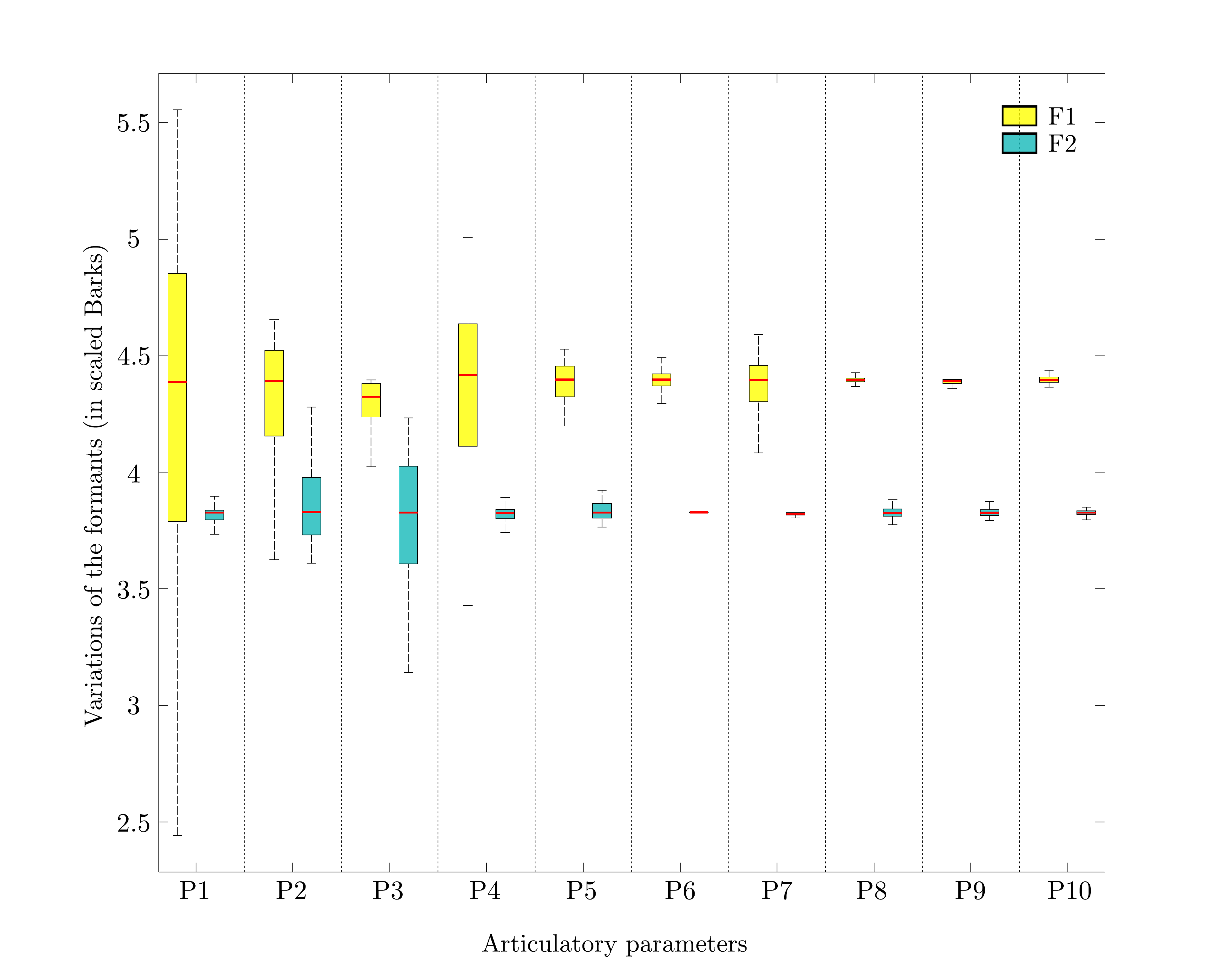}
\caption{\label{formantsVariations} Variations of the formant values (y-axis) implied by movements of each articulatory parameter (x-axis) uniformly sampled in the interval $[-1, 1]$, with $1000$ samples per parameter. When a given articulator varies, the others are kept in their rest position. %\todo{@Jules: can you please indicate the values of X and N? Thx}
}
\end{figure}

Our interpretation of why $P1$ (mainly corresponding to a jaw movement, see \figurename~\ref{fig:art_diva}) is the most frequent degree of freedom to be freed first is that:
\begin{itemize}
    \item The jaw position has a strong influence on the $F1$ frequency, especially around the articulatory rest position from which the optimization process starts. This effect is shown on \figurename~\ref{formantsVariations}.
    \item $F1$ has an important influence on auditory perception \cite{Schroeder1979}, as we modeled using a scaling factor, and thus on the cost to be optimized (see the following section for a detailed analysis).
\end{itemize}

Therefore, $P1$ is the most useful articulator to achieve a variety of vowel goals in our model, because it covers the wider range of $F1$ values (see \figurename~\ref{formantsVariations}), which has a strong influence on the cost function because of its particular weighting (three times more than $F2$ in our model). It is important to note that the result of \figurename~\ref{globalRecrut} strongly depends on this particular weighting, which is based on psychoacoustic considerations \cite{Schroeder1979,Schwartz1997}. Therefore, our model shows that the strong effect attributed to $F1$ in psychoacoustics can result in the predominance of the articulator principally controlling it (i.e. $P1$ in our model, related to jaw movements) as a way to efficiently achieve a wide variety of auditory goals. This point will be analyzed in more detail in the next experiment.

These results provide an original interpretation for explaining the predominant role of the jaw in human speech evolution and acquisition. As we discussed in Section~\ref{sec_related_work_vocal}, it has been proposed that this particular role could be due to evolutionary precursor behaviors involving jaw cycling such as mastication and ingestion~\cite{macneilage98}, as well as non-human primates communicative gestures such as lipsmacks and tonguesmacks~\cite{ghazanfar2012cineradiography}. Here we show that pure learning mechanisms, driven by stochastic optimization within a space constrained by vocal morphological properties, can also be involved due to the particular role of the jaw and the first formant in vowel production and perception.

The next section provides a sensitivity analysis emphasizing how the stochastic optimization process allows a dynamical ``freezing'' and ``freeing'' of the different articulators according to their impact on the cost function minimization, and in particular analyzing in more detail the relative impact of jaw exploration during the course of vocal optimization. 

\subsection{Experiment 3: Sensitivity Analysis}

% \todo{This section is difficult to understand. If possible, it should be rewritten as follows:
% \begin{itemize}
%   \item 2 sentences explaining what was done, and why.
%   \item description of the experiments
%   \item presentation of the results
% \end{itemize}
% }

To better understand how the stochastic optimization process structures exploration by automatically freeing and freezing degrees of freedom, this experiment performs a sensitivity analysis quantifying the respective influence of articulatory parameters on the cost function during the realization of a particular vowel. %This shows that stochastic optimization recruits in priority articulators that have a greater influence on the cost function, providing a more detailed understanding of how such a mechanism can explain the predominance of jaw movements in early vocal development.

\subsubsection{Method}

We analyze the respective influence on the cost function of the articulatory parameters during the course of an optimization run to achieve a particular vowel. The aim is to show that stochastic optimization recruits in priority articulators that have a greater influence on the cost function, providing a more detailed understanding of how such a mechanism can explain the predominance of jaw movements in early vocal development.

The optimization is run on the seven articulatory parameters with the goal of achieving the vowel \vowel{u}. 
%\figurename~\ref{fig:results} shows different recruitment orders according to the vowel the optimization process is learning.
%To understand this process in more detail, we relate the dynamical ``freezing'' and ``freeing'' of articulators to their relative influence on the cost function. 
However, for the sake of simplicity, we restrict our analysis to the role of $P1$ (mainly controlling the jaw position) and $P3$ (mainly controlling the tongue front/back position) on the cost function.
We chose these articulators because $P1$ mainly influences $F1$ values whereas $P3$ mainly influences $F2$ ones (as shown in \figurename \ref{formantsVariations}), allowing to better distinguish their respective roles in achieving the vowel \vowel{u}.
%Then we compute the partial derivatives of the cost function
The gradient of the cost function with respect to $P1$ and $P3$ is computed on regularly-sampled $(P1,P3)$ tuples. The partial derivatives are computed in an empirical manner using Equation~(\ref{Jdiff}), by looking at the cost function variations induced by small variations of the articulatory parameters around the sampled $(P1,P3)$ values. The values of the cost function $J$ in Equation~(\ref{Jdiff}) are computed by fixing the five other articulators to values allowing the production of the target vowel \vowel{u}, where we use the values obtained at the end of the optimization run.

\begin{equation}
\frac{dJ}{dPi}(Pi) = \frac{J(Pi + dPi) - J(Pi - dPi)}{2\times dPi}%, Pi\in \{P1, P3\}
\label{Jdiff}
\end{equation}
%However the optimization is run on the seven articulatory parameters in order to allow, the other ones being fixed to a position allowing the production of the vowel \vowel{u}. The partial derivatives are computed in an empirical manner, by looking at the cost function variations induced by small variations of the articulatory parameters around the current position.

This allows the quantification of the influence of $P1$ and $P3$ on the cost function depending on their values. % at dif at various moments within the optimization process, depending on what is the current best vocal solution for a given target. %, to assess if \PIBB is mainly recruiting the parameter displaying the greater influence. If it is confirmed, this will support our claim that $P1$ is n general recruited first because of its particular influence on $F1$. 
Then, we define the relative influence on the cost function of $P1$ over $P3$ by the ratio:
\begin{equation}
ratio(P1,P3) = log_{10}\left(\frac{|\frac{dJ}{dP1}(P1)|}{|\frac{dJ}{dP3}(P3)|}\right).
\label{Jratio}
\end{equation}

Therefore $ratio(P1, P3)$ is the logarithm of the ratio between the absolute partial derivatives with respect to $P1$ and $P3$ (the partial derivatives being defined in Equation~\ref{Jdiff}). It is positive when $P1$ has a greater influence than $P3$ on the cost function, negative otherwise.

\subsubsection{Results}

\begin{figure*}[!t]
\centering
\subfloat[Update 1]{\includegraphics[width=2in]{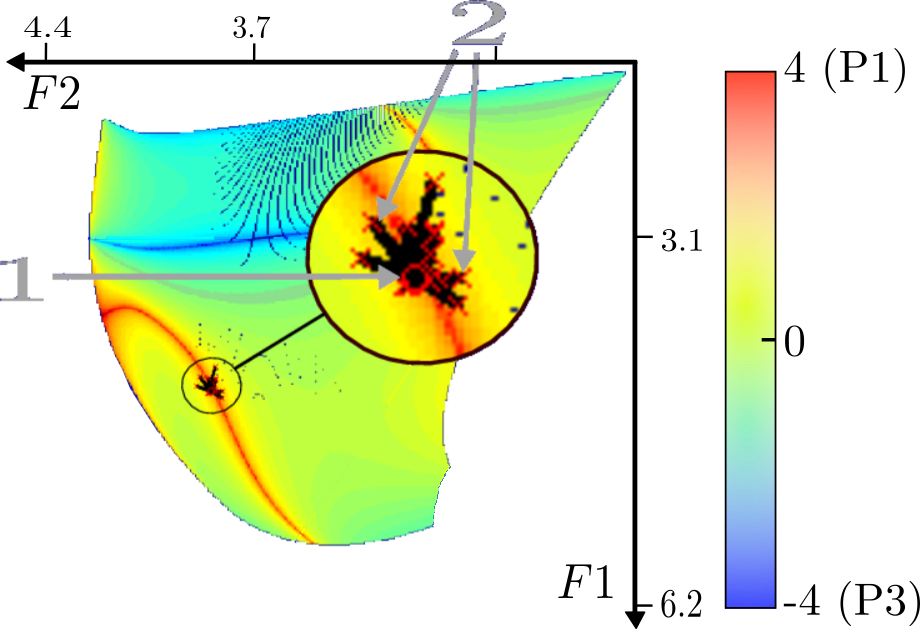} \label{simuSnapshots_first}}
\hfill
\subfloat[Update 5]{\includegraphics[width=2in]{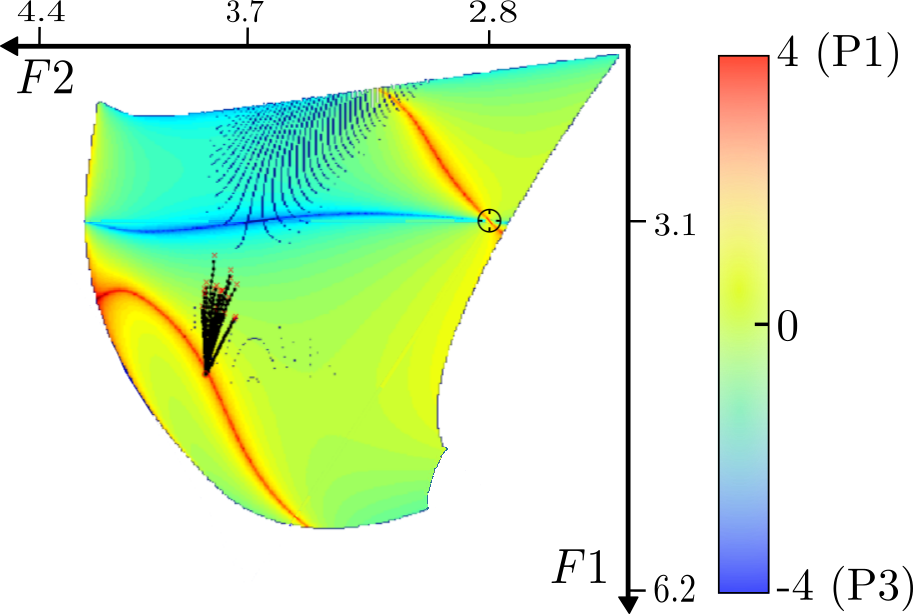}}
\hfill
\subfloat[Update 9]{\includegraphics[width=2in]{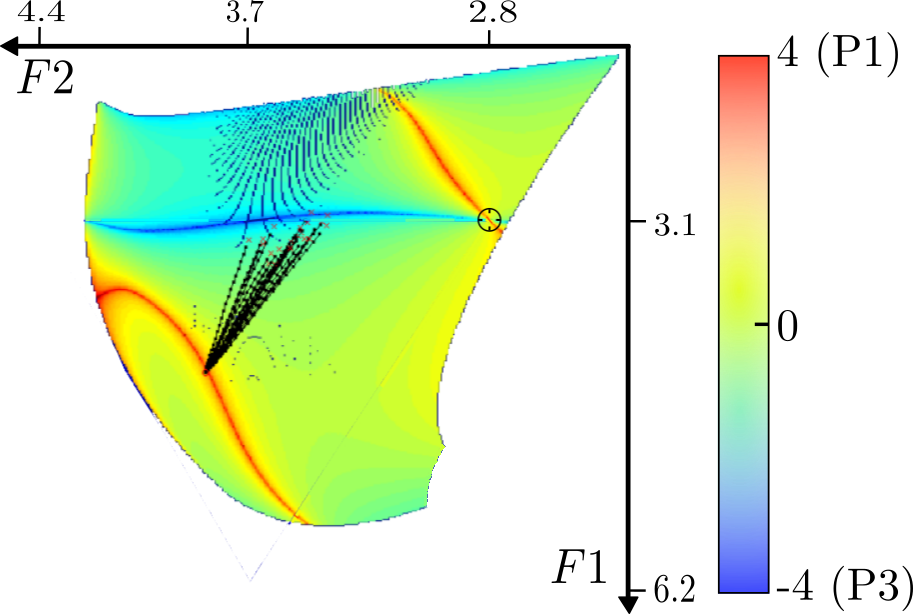}}
\vfill
\subfloat[Update 13]{\includegraphics[width=2in]{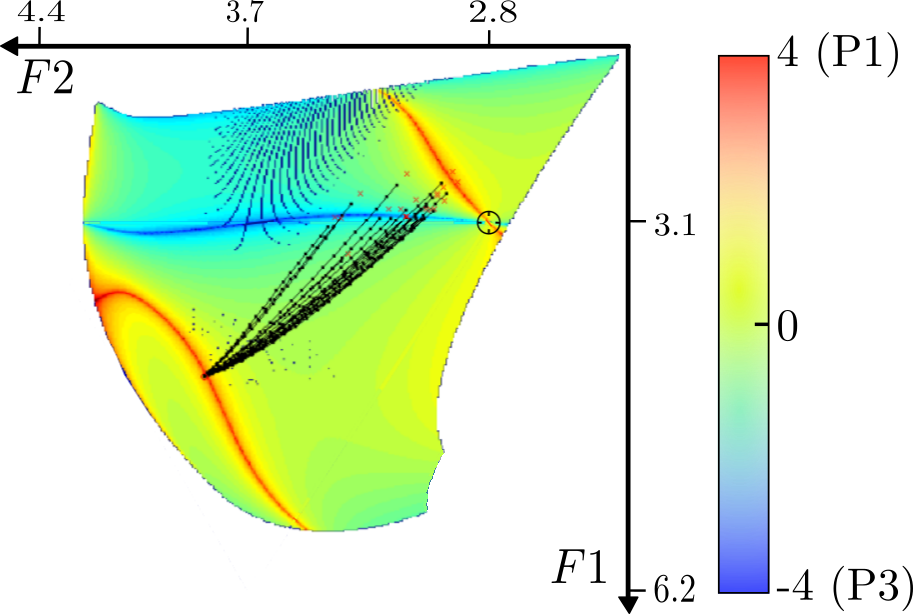}}
\hfill
\subfloat[Update 17]{\includegraphics[width=2in]{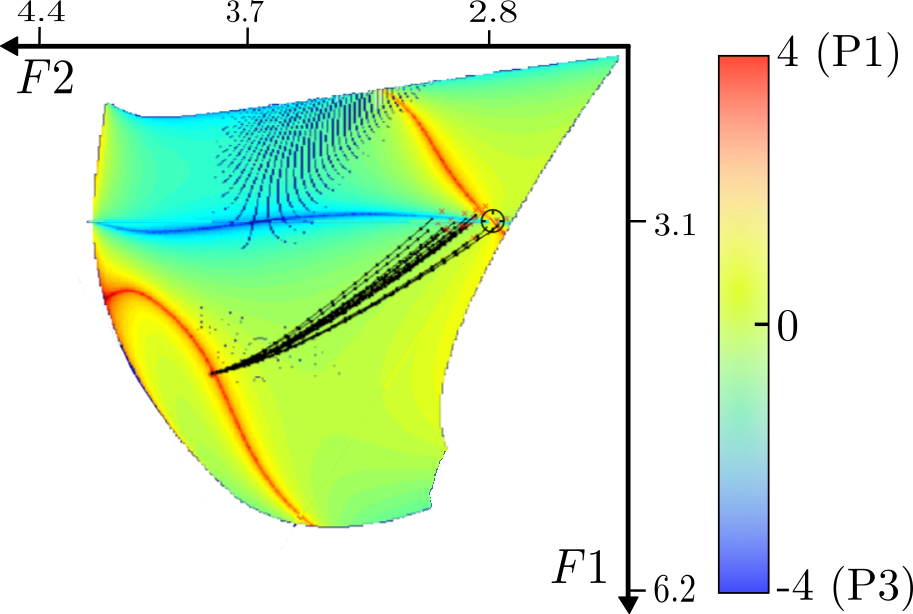}\label{simuSnapshots_beforelast}}
\hfill
\subfloat[Update 21]{\includegraphics[width=2in]{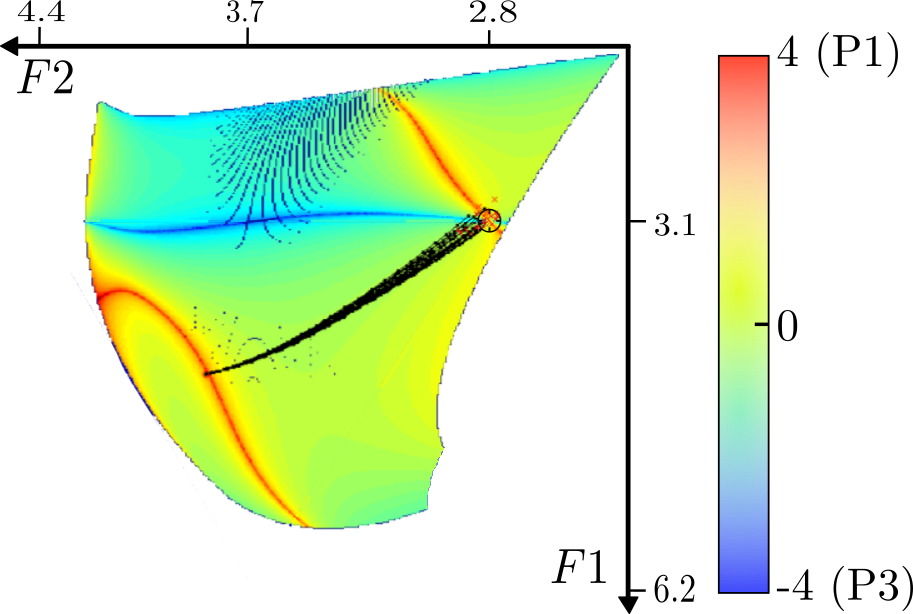} \label{simuSnapshots_last}}
\vfill
\subfloat[Relative exploration values where the update number of the snapshots are signalized by the orange arrows.]{\includegraphics[width=3in]{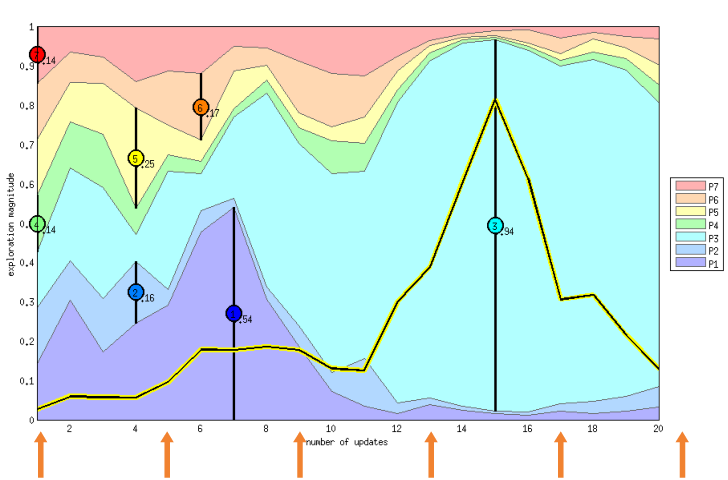} \label{snapshot_exploration}}

\caption{Snapshots from a \PIBB exploration to learn to  produce the vowel \vowel{u} (the small circle at coordinates $(F1=3.1, F2=2.8)$).
  Sub-figures \ref{simuSnapshots_first} to \ref{simuSnapshots_last} show the deployment of the $K$ concurrent samples at particular updates of the optimization process. The black lines represent the formant's trajectories, they all start from the same position (1 in sub-figure \ref{simuSnapshots_first}) which is produced with all the articulators at their rest position. The end of the trajectories are represented by a red cross (2 in sub-figure \ref{simuSnapshots_first}) and used to compute the costs. Superimposed to these trajectories is a heatmap, identical in all the sub-figures, representing the relative influence on the cost function of $P1$ over $P3$ as computed by Equation~(\ref{Jratio}). For each tuple of $(P1,P3)$ values regularly sampled in the range $[-2.5,2.5]^2$ with steps of $0.01$, we first compute the resulting formant values $(F1,F2)$ returned by the synthesizer, fixing the five other articulatory parameters to a constant value allowing the production of the target vowel \vowel{u}. Then we compute the ratio given by  Equation~(\ref{Jratio}) for each of the $(P1,P3)$ tuples. The heatmap shows the value of this ratio for each $(P1,P3)$ tuple and plotted at their resulting positions in the $F1-F2$ plane. In red areas $P1$ has a greater influence on $J$ than $P3$, while blue areas correspond to the reverse situation and green to neutral areas where both parameters have similar contribution. %This map is computed from equation \ref{Jratio} and mapped to the formant space with the synthesizer. To do so we fixed all the other parameter to their final values (after the optimisation), we varied $P1$ and $P3$ and we colored the point of the formant space corresponding to each parameter combination.  
  Sub-figure \ref{snapshot_exploration} shows the exploration magnitudes of each articulator during the optimization process, using the same conventions as in previous similar figures. The orange arrows correspond to update index of the sub-figures \ref{simuSnapshots_first} to \ref{simuSnapshots_beforelast}.}

\label{simuSnapshots}
\end{figure*}

\figurename \ref{simuSnapshots_first} to \ref{simuSnapshots_last} show the respective influence of $P1$ and $P3$ at regular time steps during a particular simulation, as computed by Equation~(\ref{Jratio}). The cost function is computed with the vowel \vowel{u} as the auditory goal. %Note that eventhough the heat maps of these figures are computed with 5 fixed parameters they provide a good overview of the cost sensitivity during the updates. 
We observe that the starting neutral configuration is in a region where $P1$ has more influence than $P3$ on the cost function, i.e. where the ratio (Equation~(\ref{Jratio})) has a high value. 

\figurename \ref{snapshot_exploration} 
shows that exploration mainly appears on $P1$ until the $9^{th}$ update. This is explained by the greater influence of $P1$ on the cost function at the beginning of the optimization process, i.e. around the rest position and in early vocal trials achieving small perturbations of this initial position, as observed on \figurename \ref{simuSnapshots_first} where the red color indicates that perturbations of $P1$ have a higher influence on the cost function. As the learning progresses towards the goal (end position of the trajectories getting closer to \vowel{u}), the optimization process reaches a blue ridge (Update 9): from this state, perturbations of $P3$ therefore display a greater influence on the cost function than perturbations of $P1$ and this explains the freeing of $P3$ and the freezing of $P1$ as observed on \figurename \ref{snapshot_exploration}.

\subsubsection{Discussion}

%These results are therefore coherent with our claim that a larger exploration magnitude of an articulator corresponds to a higher influence on the cost function of that articulator.
%$P1$ indeed both displays a higher influence on the cost function and to a larger exploration magnitude from the $1^{st}$ to the $9^{th}$ update. From this latter, the vocal tract is in a in configuration displaying a much greater influence of $P3$ rather than $P1$ on the cost function (in the blue ridge corresponding to a low ratio). Supporting our hypothesis again, we observe on Figure~\ref{snapshot_exploration} %\todo{@Jules: same here} a predominant exploration on $P3$ from the $9^{th}$ update to the end of the simulation.

%\todo{TO BE REMOVED IN THE CAPTION: We observe that around the rest position, and in early vocal trials achieving small perturbations of this initial position, the red color indicates that perturbations of $P1$ produce a higher influence on the cost function (thus triggering a higher relative exploration magnitude as observed in subfigure \ref{snapshot_exploration}). As the learning progresses towards the goal (end position of the trajectories getting closer to \vowel{u}), the optimization process reaches a blue ridge (Update 9): from this state, perturbations of $P3$ become more impactful and this explains the freeing of $P3$ as observed in previous graphs. }

This sensitivity analysis provides evidence that the stochastic optimization process will favor the recruitment of articulators having greater influence on the cost function minimization at a given update. In other words, exploration takes place on articulators which are the most useful to move closer to the auditory goal according to the current configuration of the vocal tract. Since $P1$ (mainly controlling jaw movements) has a greater influence on $F1$ (\figurename \ref{formantsVariations}) and $F1$ has a higher weight in the cost function computation (Eq.~\ref{J}), this analysis provides a computational understanding of the results obtained in \figurename~\ref{globalRecrut}, where we observed a strong tendency of recruiting in priority $P1$ to achieve various vowel goals. 

\section{Conclusion}

It has been proposed that canonical babbling is so robust that a number a forces probably act on it \cite{Oller2000}. Traditional propositions of these forces concern the derivation of prelinguistic behaviors in the course of human evolution, for example non-human primate orofacial communicative gestures \cite{macneilage98}; the influence of the social environment \cite{Warlaumont_2013_NN}; or the role of curiosity-driven learning in exploring the articulatory-auditory space \cite{Moulin-Frier_Frontiers_2013}. This paper proposes an original hypothesis regarding the predominant role of the jaw in infant vocal development, where we suggest that it could be a result of stochastic learning processes allowing the production of various auditory effects. This resulting predominance of jaw movements would then favor the emergence of canonical babbling in infant vocal development. This work takes inspiration of a previous model showing that such exploration strategies implied by a stochastic optimization process provides a cognitive reason for the proximo-distal law of arm development \cite{stulp12emergent,stulp13adaptive}.

For this aim, we have developed a computational model using an articulatory synthesizer, movement generation and auditory perception processes, coupled with the stochastic optimization algorithm \PIBB. We have run simulations where the system iteratively optimizes the reaching of various auditory vowel goals and we have performed various analyses on the underlying results: extraction of exploration magnitude and sensitivity analysis during the exploration process, as well as the order of articulator recruitment. These results show that the order of recruitment was dependent of the auditory goal to be reached and that, on average on various auditory goals, the jaw is predominantly and firstly recruited by the optimization process. Moreover, a sensitivity analysis suggests that the order of recruitment is determined by the relative influence of each articulator on the cost function at a given time step.

Our hypothesis seems simpler than the aforementioned previous propositions in the sense that it relies on a general optimization process allowing to find adequate motor commands in order to reach various sensory goals. We do not claim that other factors, such as prelinguistic behaviors, the social environment, or curiosity-driven learning, do not play a role in this process. Actually, social guidance and curiosity-driven learning could both be mechanisms accounting for the selection of auditory goals that are set by the experimenter in the model presented in this paper. 
As a whole, we agree with Oller's proposition (\cite{Oller2000}) that a number of forces probably act in favor of canonical babbling, thus explaining its robustness. Here we have proposed an original one which has the advantage of relying only on very general learning mechanism and also generalizes on other developmental aspect such as the proximodistal law of arm control.

\section*{Acknowledgment}
This work was partially financed by ERC Starting Grant EXPLORERS 240 007.

The authors would like to thank Louis-Jean Bo\"{e} for the design of \figurename~\ref{fig:trans_artic_acoust} (vocal tract by Sophie Jacopin).

\newpage

\bibliographystyle{abbrv}
%\bibliographystyle{IEEEtran}
\bibliography{main}
%bibliography{../bibliography}
%\bibliography{../bibliography}

\end{document}